\documentclass[iop,revtex4]{emulateapj}

\usepackage{hyperref} 

\begin{document}
\slugcomment{Accepted for publication in ApJ}

\shorttitle{ISM in the \emph{Kepler} Field}
\shortauthors{Johnson et al.}

\title{The Interstellar Medium in the \emph{Kepler} Search Volume}

\author{Marshall C. Johnson\altaffilmark{1}, Seth Redfield\altaffilmark{2}, and Adam G. Jensen\altaffilmark{3}}

\altaffiltext{1}{Department of Astronomy, University of Texas at Austin, 2515 Speedway, Stop C1400, Austin, TX 78712, USA; mjohnson@astro.as.utexas.edu}
\altaffiltext{2}{Astronomy Department, Van Vleck Observatory, Wesleyan University, Middletown, CT 06459, USA}
\altaffiltext{3}{Department of Physics and Physical Science, University of Nebraska-Kearney, Bruner Hall of Science, 2401 11th Ave, Kearney, NE 68849, USA}

\begin{abstract}
The properties of the interstellar medium (ISM) surrounding a planetary system can impact planetary climate through a number of mechanisms, including changing the size of the astrosphere (one of the major shields for cosmic rays) as well as direct deposition of material into planetary atmospheres. In order to constrain the ambient ISM conditions for exoplanetary systems, we present observations of interstellar Na~\textsc{i} and K~\textsc{i} absorption towards seventeen early-type stars in the \emph{Kepler} prime mission field of view. We identify 39 Na~\textsc{i} and 8 K~\textsc{i} velocity components, and attribute these to eleven ISM clouds. Six of these are detected towards more than one star, and for these clouds we put limits on the cloud properties, including distance and hydrogen number density. We identify one cloud with significant ($\gtrsim1.5$ cm$^{-3}$) hydrogen number density located within the nominal $\sim$100 pc boundary of the Local Bubble. We identify systems with confirmed planets within the {\it Kepler} field of view that could lie within these ISM clouds, and estimate upper limits on the astrosphere sizes of these systems under the assumption that they do lie within these clouds. Under this condition, the Kepler-20, 42, and 445 multiplanet systems could have compressed astrospheres much smaller than the present-day heliosphere. Among the known habitable zone planet hosts, Kepler-186 could have an astrosphere somewhat smaller than the heliosphere, while Kepler-437 and KOI-4427 could have astrospheres much larger than the heliosphere. The thick disk star Kepler-444 may have an astrosphere just a few AU in radius.
\end{abstract}


\section{Introduction}
\label{introduction}

The interstellar medium (ISM) is often studied by observing absorption lines in stellar spectra, which have been imprinted by intervening interstellar matter \citep[e.g.,][]{Hartmann04}. For ground-based optical spectra this can most easily be accomplished by observing early-type stars, which have spectra largely free of stellar absorption lines; any stellar lines that are present are typically significantly rotationally broadened. Interstellar matter imprints a signature of narrow absorption lines on the stellar spectrum. High resolution spectra allow the separation of multiple velocity components (corresponding to different ISM clouds), typically separated by a few km s$^{-1}$.
The primary ions accessible in the optical are Na~\textsc{i}, K~\textsc{i}, and Ca~\textsc{ii}, while many other species are observable in the ultraviolet \citep[e.g.,][]{RedfieldLinsky04}.
The distance to the star provides an upper limit on the distance to the interstellar cloud, while the most distant stars in the same region of the sky that show no absorption provide a lower limit to this distance \citep[e.g.,][]{Lallement,Peek11}. This technique has been used to map the three-dimensional structure of the ISM on both large and small scales. For instance, \cite{Lallement} mapped the ISM in all directions out to distances of $\sim250$ pc, while \cite{RedfieldLinsky08} mapped the structure and kinematics of the local ISM (LISM), the collection of clouds in the immediate vicinity of the solar system. 
Most previous work using these techniques covered much or all of the sky \citep[e.g.,][]{WeltyHobbs01,Lallement}. Targeted surveys of a few selected fields have been conducted, such as towards the Hyades \citep{RedfieldHyades} and in the direction of the solar system's past trajectory \citep{WymanRedfield}. In this paper we present the first targeted survey of the ISM in a direction of exoplanetary interest: the {\it Kepler} field of view.

In order to provide motivation and background for this work, in \S\ref{helioISM} we introduce the interactions between the heliosphere (and by extension astrospheres) and the ISM, while in \S\ref{ISMeffects} we review the effects of variable astrosphere size upon any planets residing within. In \S\ref{relating} we discuss how our observables in terms of the ISM and exoplanetary properties can be used to constrain astrospheric properties, and provide a brief introduction to the {\it Kepler} mission.

\subsection{The Heliosphere and the ISM}
\label{helioISM}

In our own solar system, the solar wind carves out a cavity in the surrounding ISM known as the heliosphere. As the solar wind moves outward, it eventually encounters the ISM, resulting in a series of shocks and boundaries. The innermost of these is the termination shock, where the solar wind decelerates from supersonic to subsonic velocities. Both Voyager 1 and Voyager 2 have crossed this shock, at distances of 94 and 84 AU, respectively \citep{Stone05,Stone08}. The corresponding outer boundary where the ISM becomes subsonic, the bow shock, is expected to lie at a distance of $\sim250$ AU \citep{Izmodenov05}; however, recent work has suggested that the solar system is moving through the LISM at an insufficient velocity to induce the formation of a bow shock \citep{McComas12}. In between these two shocks lies the heliopause, the contact interface between the solar wind and the LISM. Voyager 1 crossed the heliopause at a distance of 121 AU \citep{Gurnett13}, while Voyager 2 has yet to cross. The turbulent region between the termination shock and the heliopause is known as the heliosheath. Other stars also have stellar winds, and so will possess analogues to the heliosphere, known as astrospheres, with analogous structures (i.e., termination shock, astropause, bow shock, astrosheath).

Planetary surfaces can be protected from cosmic rays by three major shields--the interplanetary magnetic field carried by the solar (stellar) wind within the heliosphere (astrosphere), the planetary magnetic field, and the planetary atmosphere \citep{Scherer06}. Because the solar (stellar) wind carries its embedded magnetic field as far as the heliopause (astropause), when we refer to the ``heliosphere size'' or ``astrosphere size'' for the remainder of this article, we are referring specifically to the distance from the star to the heliopause or astropause. We note that the heliosphere and astrospheres are not spherical, but are rather swept back into a ``cometary'' shape by the star's motion through the ISM; the sizes that we quote are the distances to the astropause in the upwind direction. The importance of the heliopause is illustrated by the fact that the cosmic ray flux measured by Voyager 1 increased by $\sim9$\% in less than a day when it crossed the heliopause \citep{Krimigis13}. The modulation of cosmic rays within the heliosphere is caused primarily by cosmic rays scattering through magnetic fields in the turbulent solar wind and the heliosheath \citep{Muller06}. Thus, the efficacy of the heliosphere as a cosmic ray shield will be affected by both the size of the heliosphere and the degree of turbulence. The turbulence is caused in large part by the scattering of pickup ions within the heliosphere \citep[e.g.,][]{Isenberg03,Gamayunov12}, which is in turn related to the neutral density in the surrounding ISM.

The size of the astrosphere surrounding a star is determined by a momentum balance between the outgoing stellar wind and the streaming ISM. Higher ISM momentum--i.e., higher ISM density or higher velocity of the star relative to the ISM--will reduce the size of the astrosphere. If the size of the astrosphere is reduced, the distance over which cosmic rays cross magnetic field lines is reduced, and thus the shield is less effective; the cosmic ray flux at any planets will increase. Quantitatively, \cite{Muller06} found that during a passage through a dense interstellar cloud where the heliopause lies at a distance of 26 AU from the sun, the intensity of cosmic rays with energies between 300 MeV and 1 GeV at the Earth will increase by a factor of 1.4 to 7.6 over the present flux, depending upon assumptions about the power spectrum of the solar wind turbulence, corresponding to the relative importance of cosmic ray modulation in the solar wind versus in the heliosheath.

The ISM in the immediate vicinity of the solar system, the LISM, has been extensively studied; see \cite{ARAA} for a review of this work. The Sun is currently near the edge of the Local Interstellar Cloud (LIC). Observations by \cite{Linsky00} placed a lower limit on the neutral hydrogen number density of the LIC of $n_{\mathrm{H I}}\sim0.10$ cm$^{-3}$; this may be regarded as a firm lower limit for the total column density, as H~\textsc{i} will be only one constituent of the LIC. The combination of more detailed models and observations find total hydrogen number densities ranging from 0.23 to 0.30 cm$^{-3}$ \citep{SlavinFrisch08,Heerikhuisen14}. We will thus adopt 0.2 cm$^{-3}$ as the LIC density for the remainder of this work.

\subsection{Effects of the ISM upon Planets}
\label{ISMeffects}

The motion of planet-hosting stars through the ISM can have several effects upon the planets, due to the changing astrosphere size (which affects the cosmic ray flux experienced by the planet) as well as direct deposition of interstellar material onto the planetary atmosphere. We will briefly summarize these processes. 

The cosmic ray flux can have a number of effects on the climate and habitability of planets \citep[see][for a review of these processes]{Scherer06}. Cosmic rays have been tentatively linked to cloud nucleation \citep[e.g.,][]{SvensmarkFriisChristensen97,MarshSvensmark00,Kirkby11}, although this is controversial \citep[e.g.,][]{Kernthaler99,Laken13}. \cite{ErlykinWolfendale11} surveyed the literature and concluded that the contribution of cosmic rays upon the Earth's tropospheric cloud cover is of order 1\%. Lightning may also be affected by cosmic rays, as cosmic ray-induced atmospheric ionization could affect the global electric circuit \citep{Chronis09}. Both clouds and lightning can have broader effects on climate; clouds change the planet's albedo, thus affecting surface temperatures, while lightning produces NO$_{\mathrm{x}}$ compounds in the troposphere, which are precursors to the formation of ozone, a greenhouse gas \citep{Allen10}. 

More extreme effects will occur if two of the Earth's cosmic ray shields are lost simultaneously--that is, if the Earth's magnetic field undergoes a polarity reversal when the solar system is traversing a dense interstellar cloud. This scenario was considered by \cite{Pavlov05O3}. During such a reversal, the overall strength of the Earth's magnetic field decreases. The resulting increase in cosmic rays causes ionization of atmospheric molecules, which directly forms NO$_{\mathrm{x}}$ compounds. Unlike in the troposphere, stratospheric ozone tends to be {\it destroyed} by NO$_{\mathrm{x}}$. \cite{Pavlov05O3} found that this could decrease ozone concentrations by 40\% or more globally. For life-bearing planets, the cosmic ray flux has a direct bearing on mutation rates and further radiation damage, both directly through impacts of cosmic rays and secondary particles from cosmic ray air showers on living tissue, and indirectly through reduction of the global ozone layer \citep{Pavlov05O3}, resulting in higher doses of ultraviolet radiation. They found that, given the duration of such cloud crossings and the frequency of magnetic field reversals, a reversal should occur at least once during each cloud crossing.

Interstellar dust can also affect the climate of planets. If sufficient amounts of dust are swept up into a planetary atmosphere, it can cause a ``reverse greenhouse'' effect, radiating efficiently in the infrared and cooling the planet \citep{Pavlov05}.

The most extreme possible effects will occur if the ISM density is sufficiently high that the astrosphere retreats inside the habitable zone. This possibility was investigated by \cite{SmithScalo}, who referred to this occurrence as ``descreening.'' They found that for a Sun-like star an ambient ISM density of $\sim$600 cm$^{-3}$ is necessary for descreening to occur. Clouds with such high densities are small, and they calculated a rate of these events of $1-10$ Gyr$^{-1}$ for sun-like stars, and less frequently for later-type stars due to the greater proximity of the habitable zone to these stars. \cite{YeghikyanFahr04} found that if the Earth is directly exposed to the ISM, the increased flux of hydrogen could alter the atmospheric chemistry, resulting in severe ozone depletion. This could also result in an ice age.

\subsection{Relating Observables to Astrospheres}
\label{relating}

The {\it Kepler} mission searched for planets down to the size of the Earth using the transit method, by making nearly continuous photometric observations of nearly 150,000 stars for more than four years and searching for the periodic flux decrements associated with transiting planets \citep[e.g.,][]{Seader15}. The region that {\it Kepler} surveyed for planets (hereafter denoted the ``{\it Kepler} field of view'' or the ``{\it Kepler} search volume'') is centered at an RA of $19^h 22^m 40^s$ and a Declination of $44^{\circ} 30' 00''$, and fits within a square $13.9^{\circ}$ on a side, although the filling factor of the {\it Kepler} CCDs within this square is less than unity \citep{KepFOV}; the {\it Kepler} CCDs cover an area of 115 square degrees \citep{Seader15}. {\it Kepler} is the first mission capable of detecting potentially habitable planets--i.e., those with masses close to that of the Earth and orbiting far enough from their host stars that they could possess surface liquid water--where the effects of interactions with the ISM described above could have interesting consequences. {\it Kepler} has already found a number of small planets in the habitable zone \citep[e.g.,][]{Torres15}. We thus chose this region of the sky for a survey to determine the structure of the ISM, with the objective of constraining the interstellar environment for the known potentially habitable planets discovered by {\it Kepler}. 

The first to consider the astrosphere sizes of exoplanetary systems was \cite{Frisch93}, who calculated the astrosphere size for a sample of 60 nearby G-type stars. At the time, however, none of these stars was known to possess planets. With the multitude of exoplanets now known, in particular the large number of candidate planets from \emph{Kepler}, including a number of confirmed habitable zone super-Earths, we can today make a much more specific study of the ambient ISM properties for known planetary systems. To our knowledge the present work is the first to address the ISM environments of known exoplanetary systems. In analogy to many recently-coined terms referring to extrasolar equivalents to features of the solar system, we propose the term \emph{exoLISM} to refer to the ISM in the vicinity of exoplanetary systems. In complementary work, Edelman et al.\ (2015, in prep) have searched for Ly $\alpha$ absorption from the hydrogen walls of astrospheres associated with nearby planet-hosting systems in order to estimate the mass loss rates \citep[i.e., the stellar wind properties;][]{Wood05} of these stars. Most of the {\it Kepler} systems, however, are too distant for this kind of observations, as absorption from intervening interstellar gas overwhelms the weak astrospheric signal. Therefore, mapping the ISM in the {\it Kepler} search volume is currently the best option for constraining the astrosphere sizes for {\it Kepler} planets.

We note that our techniques are only capable of probing two of the four parameters most responsible for determining the size of an astrosphere ($\rho_{ISM}$, $v_{ISM}$, $\rho_{SW}$, $v_{SW}$, where ``SW'' denotes the stellar wind). We can measure the ISM density and the ISM velocity, the latter of which is necessary to determine the relative velocity between the ISM and the star. Our methods, however, are not capable of probing the transverse velocities, and hence the full 3-D space motions, of ISM clouds, and so we will have to estimate the full 3-D relative velocity from the difference in radial velocities between the star and the ISM. In this work we do not measure the other two parameters upon which the astrosphere size is most strongly dependent, namely the stellar wind velocity and density; these can be measured for nearby stars through astrospheric absorption, as described above. The astrosphere size is also dependent upon the other properties of the ISM, such as the interstellar magnetic field and ionization properties \citep{Zank13}, but we are also unable to determine these parameters from our data.

This work has a number of secondary applications in addition to estimation of astrosphere sizes. First, the distances to many of the \emph{Kepler} candidate systems and even confirmed planetary systems are rather uncertain. Determination of the locations and distances to ISM clouds within the \emph{Kepler} field of view can help constrain the distances to these systems; if absorption from a given cloud is seen in the spectrum of a planet host star, then that star is at least as distant as that cloud. Conversely, if a star is located in the same part of the \emph{Kepler} field of view as a cloud but does not show interstellar absorption in its spectrum, then the star is at most as distant as that cloud. Second, this work will allow the prediction and modeling of interstellar lines in the spectra of \emph{Kepler} targets, potentially an important source of spectral contamination for some applications.

\section{Observations, Data Reduction, and Modeling}

Targets were selected from the updated \emph{Hipparcos} catalog of \cite{vanleeuwen07}. We selected stars in the \emph{Kepler} field of view with $B-V<0.41$ %Not sure if this is the number I actually used because my notes burned up.
in order to select early-type stars, and concentrated on bright stars ($V<7$) in order to minimize the necessary observation time; $\sim100$ stars met these criteria. From this list we observed 17 targets; these consisted of the brightest stars in the sample, plus others selected to provide roughly even sampling over the full range of possible distances, from $\sim20$ pc to $\sim1$ kpc. The emphasis on very distant stars resulted in a distribution that mostly sampled the low Galactic latitude side of the {\it Kepler} field at large distances. The properties of observed stars are listed in Table \ref{startable}. 

The data were obtained using the 2.7m Harlan J.\ Smith Telescope and Robert G.\ Tull TS21 Spectrograph \citep{Tull95} at McDonald Observatory on 2010 August $16-18$ UT. The instrumental resolving power is $R\sim240,000$. The weather varied from clear to light clouds. Two stars (HIP 96288 and 96693) were observed on both August 16 and 18 due to suboptimal signal-to-noise in the first observations. All other stars were observed on only one night. Our primary spectroscopic target was the Na~\textsc{i} D1/D2 doublet, at $\lambda\lambda$ 5896, 5890 \AA. We also observed the K~\textsc{i} line at 7699 \AA. Due to the instrumental setup, however, we could not observe the other line of the K~\textsc{i} doublet.

Our reduction and analysis pipeline is the same as used on previous studies, such as \cite{Redfield07} and \cite{WymanRedfield}.
We performed the data reduction using standard \emph{IRAF} tasks.
We converted the velocities to the local standard of rest (LSR) frame, using the values of the solar motion measured by \cite{Schonrich10}. All velocities quoted in this work are in the LSR frame, unless noted otherwise.
After aperture extraction, we normalized the data by fitting a polynomial function using sigma-clipping to isolate the continuum. This process was straightforward due to the flat continua of the hot stars that we targeted. Additionally, since we have observed both lines of the Na~\textsc{i} doublet we can minimize the effects of systematic errors in the normalization by simultaneously fitting both lines. 

Next we removed the telluric lines in the data. We followed the same methodology as \cite{WymanRedfield}, using the forward modeling techniques of \cite{Lallement93} with a model telluric spectrum produced by the Atmospheric Transmission program \citep{Grossman89}. The region around both Na~\textsc{i} lines suffers from pervasive telluric contamination; the deepest telluric lines absorb at most $\sim40$\% of the signal in one pixel. For the D1 line, the region between $-15$~km~s$^{-1}$ and $+25$ km s$^{-1}$ is largely free of telluric contamination. For the D2 line, however, there are telluric features that are usually superposed upon the interstellar features. Again, fitting both the D1 and D2 lines simultaneously helps to minimize the effects of any imperfections in the telluric subtraction. There is no significant telluric contamination within $\sim80$ km s$^{-1}$ of the K~\textsc{i} line center.
For some targets we needed to perform a second normalization after the telluric subtraction, using a lower-order polynomial.

We produced model spectral lines using a Voigt line profile; for the Na~\textsc{i} lines we took into account the hyperfine structure of each line. We then convolved this model of the actual line profile with the spectrograph line spread function to produce a model of the observed spectral lines. We performed fits to the D1 and D2 lines individually and simultaneously. We determined the final line parameters by combining the values from these three fits. For each line component, the free parameters are the velocity $v$, the Doppler parameter $b$, and the column density $N$. 
We obtained error estimates using a Monte Carlo routine after the fits were performed. We used the F-test to determine the best-fitting number of line components. For cases where no interstellar absorption is detected, we calculated the upper limits on the column density from the root-mean-squared (RMS) scatter of the continuum in the region around the lines. For Na~\textsc{i}, we use the D2 line to calculate the limits due to its higher oscillator strength.

Some modifications of this process were necessary for HIP 96441, our coolest star (spectral type F4V), due to the presence of strong stellar Na~\textsc{i} and K~\textsc{i} lines in the spectrum. We produced an empirical model of the stellar lines by smoothing the spectrum and reflecting the resulting smoothed spectrum across the line center. These model lines were then subtracted from the data. This process removes the broad stellar lines while leaving any narrow interstellar lines intact. No interstellar absorption, however, is detected for this target, the nearest in our sample ($d=18.34$ pc). This is expected, as this target is well inside the boundaries of the Local Bubble \citep{Lallement}.

\begin{deluxetable*}{cccccccccc}
\tabletypesize{\scriptsize}
\tablecolumns{10}
\tablecaption{Target Sample \label{startable}}
\tablehead{ 
\colhead{HIP} & \colhead{HD} & \colhead{Name} & \colhead{RA (J2000.0)} & \colhead{Dec (J2000.0)} & \colhead{Spec. Type} & \colhead{$V$} & \colhead{$d$ (pc)} & \colhead{Exp. time (s)} & \colhead{SNR}  \\
\colhead{(1)} & \colhead{(2)} & \colhead{(3)} & \colhead{(4)} & \colhead{(5)} &\colhead{(6)} & \colhead{(7)} & \colhead{(8)} & \colhead{(9)} & \colhead{(10)} 
}

\startdata
96441 & 185395 & $\theta$ Cyg & 19 36 26.53 & +50 13 15.96 & F4V & 4.5 & $18.34 \pm 0.05$ & 900 & 130 \\
93408 & 177196 & 16 Lyr & 19 01 26.37 & +46 56 05.32 & A7V & 5.0 & $37.4 \pm 0.2$ & 1200 &  110\\
97165 & 186882 & $\delta$ Cyg & 19 44 58.48 & +45 07 50.92 & B9.5IV+ & 2.9 & $50.6 \pm 1.2$ & 360 & 160 \\
92822 & 175824 & \ldots & 18 54 47.12 & +48 51 33.84 & F3III & 5.8 & $55.3 \pm 0.8$ & 1200 & 66 \\
97700 & 188074 & \ldots & 19 51 19.38 & +47 22 38.05 & F2V & 6.2 & $62.5 \pm 1.6$ & 1200 & 53 \\
96286 & 184977 & \ldots & 19 34 39.86 & +48 09 52.38 & A9V & 6.8 & $86.5 \pm 2.8$ & 1200 & 47 \\
96195 & 184603 & \ldots & 19 33 36.44 & +38 45 43.10 & A3Vn & 6.6 & $129 \pm 8$ & 1200\tablenotemark{a} & 51 \\
96288 & 184875 & \ldots & 19 34 41.26 & +42 24 45.04 & A2V & 5.3 & $177 \pm 6$ & 1500\tablenotemark{b} &  54 \\
96693 & 185872 & 14 Cyg & 19 39 26.49 & +42 49 05.81 & B9III & 5.4 & $200 \pm 8$ & 1500\tablenotemark{b} & 72 \\
93210 & 176582 & V545 Lyr & 18 59 12.29 & +39 13 02.36 & B5IV & 6.4 & $292 \pm 26$ & 1500 & 82 \\
98194 & 189178 & \ldots & 19 57 13.87 & +40 22 04.17 & B5V & 5.5 & $340 \pm 22$ & 1200 & 82 \\
96491 & 185330 & \ldots & 19 36 56.65 & +38 23 01.77 & B5II-III & 6.5 & $361 \pm 40$ & 1200 & 53 \\
94481 & 180163 & $\eta$ Lyr & 19 13 45.49 & +39 08 45.48 & B2.5IV & 4.4 & $426 \pm 24$ & 900 & 130 \\
97845 & 188439 & V819 Cyg & 19 53 01.25 & +47 48 27.79 & B0.5IIIn & 6.3 & $503 \pm 71$ & 1200 & 38 \\
97634 & 187879 & V380 Cyg & 19 50 37.33 & +40 35 59.14 & B1III+ & 5.7 & $649 \pm 101$ & 1200 & 68 \\
95673 & 183362 & V558 Lyr & 19 27 36.40 & +37 56 28.31 & B3Ve & 6.3 & $725 \pm 152$ & 1500 & 43 \\
97757 & 188209 & \ldots & 19 51 59.07 & +47 01 38.42 & O9.5Ia & 5.6 & $1100 \pm 266$ & 1200 & 72 
\enddata

\tablecomments{(1) Star ID in the \emph{Hipparcos} catalog, in order of increasing distance. (2) Star ID in the Henry Draper catalog. (3) Bayer, Flamsteed, or variable star designation. (4) Right ascension, equinox J2000.0, in hours, minutes, seconds. (5) Declination, equinox J2000.0, in degrees, minutes, seconds. (6) Spectral type. (7) $V$-band magnitude. (8) Distance in parsecs, as determined from \emph{Hipparcos} parallaxes \citep{vanleeuwen07}. (9) Total exposure time for the star, divided into 2-4 exposures per star. (10) Signal-to-noise ratio per pixel in the region of the Na~\textsc{i} lines. Data for columns 1-8 from the SIMBAD database (http://simbad.u-strasbg.fr/simbad/).}

\tablenotetext{a}{A cosmic ray hit on the Na~\textsc{i} D1 line ruined one exposure; effective exposure length for analysis of the D1 line was 600 seconds.}
\tablenotetext{b}{These targets were observed on two nights due to low signal-to-noise in the first set of observations due to thin clouds. All other targets were observed on only one night.}

\end{deluxetable*}

\section{Analysis}

\begin{figure*}
\plotone{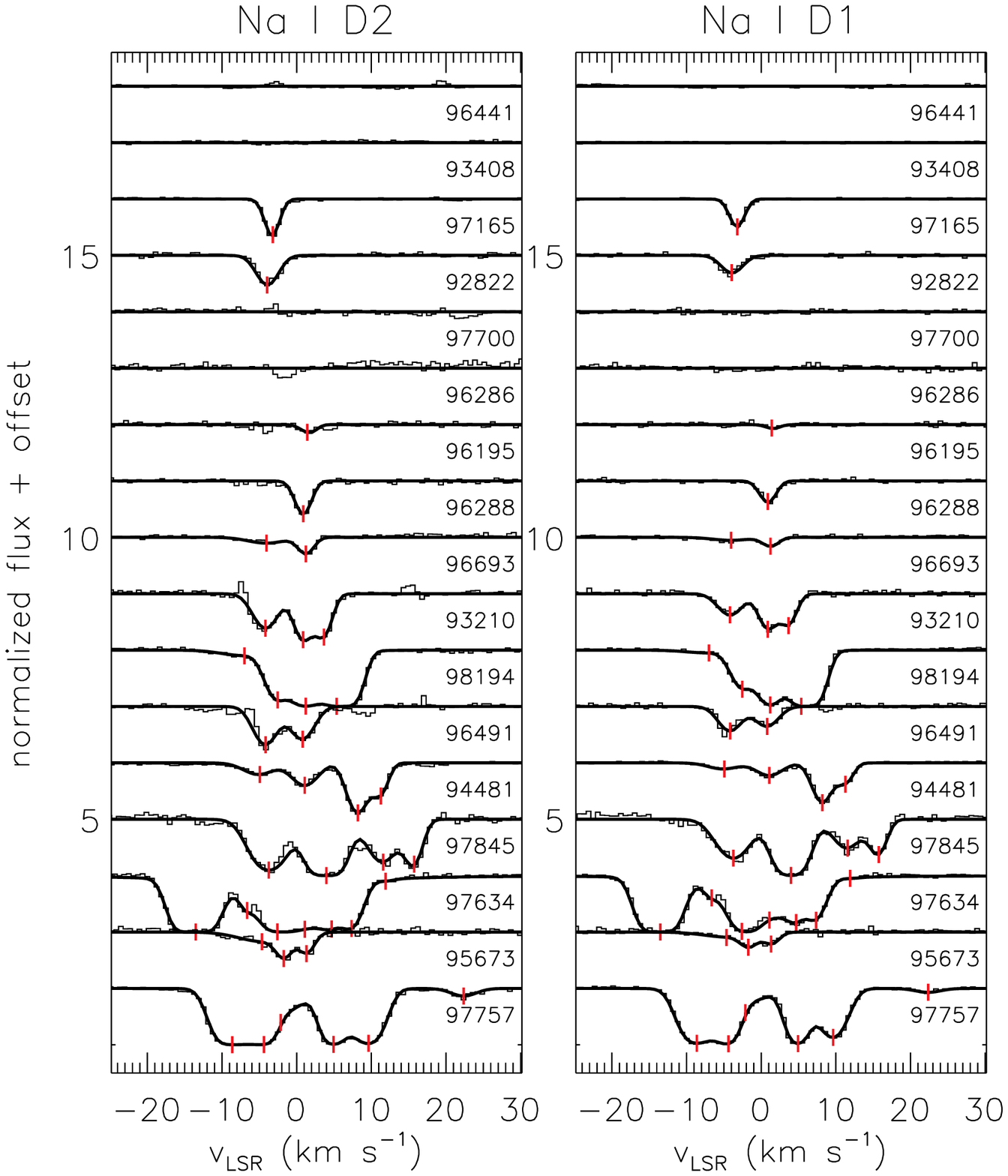}
\caption{Normalized spectra (thin lines) and fits (thick lines) at the Na~\textsc{i} D2 (left) and D1 (right) lines for all stars. Spectra are shown in order of increasing stellar distance from top to bottom, and each spectrum is offset by unity for clarity. Vertical red hatch marks denote the central velocity of each component. Each spectrum is labeled with the stellar HIP catalog number. Spikes and dips in the spectra which are not included in the fits are largely due to imperfectly-removed telluric lines. These can easily be distinguished from Na~\textsc{i} absorption because there is no corresponding feature in the  other Na~\textsc{i} line. \label{allna1}}
\end{figure*}

\begin{figure}
\plotone{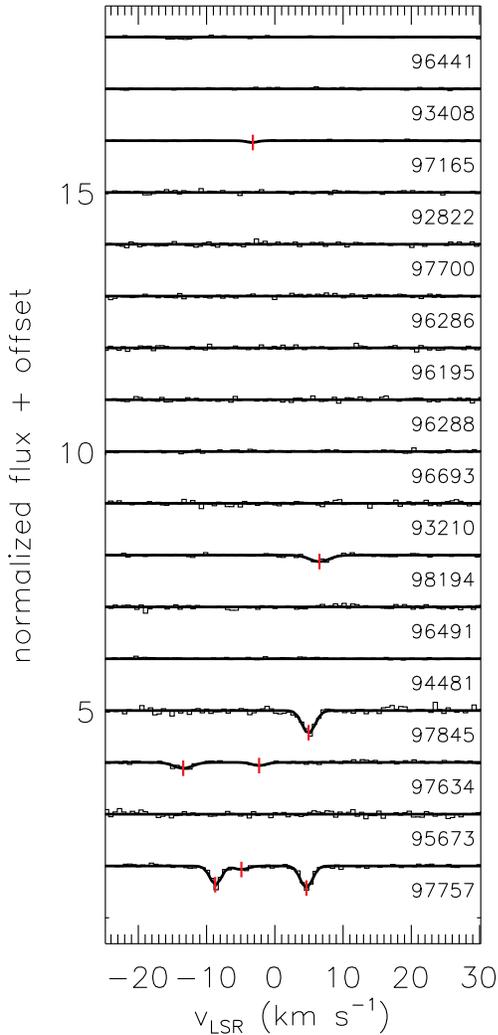}
\caption{Same as Fig.~\ref{allna1}, except for the K~\textsc{i} line. \label{allk1}}
\end{figure}

We detect Na~\textsc{i} towards 13 of the 17 targets (totaling 39 line components), and K~\textsc{i} towards 5 of the 17 (8 total line components). All K~\textsc{i} components correspond to a detected Na~\textsc{i} component to within 1.2 km s$^{-1}$. We show the data and fits in Fig.~\ref{allna1} for Na~\textsc{i} and Fig.~\ref{allk1} for K~\textsc{i}. A number of the components, especially for the more distant targets, are saturated and/or blended with other components. The number of components and the total column density generally increase with distance. Detected Na~\textsc{i} components cover a wide range of LSR velocities, from $-14$~km~s$^{-1}$ to $+22$ km s$^{-1}$; all except three components have $-10$ km s$^{-1} < v < +13$ km s$^{-1}$.

\begin{figure}
\plotone{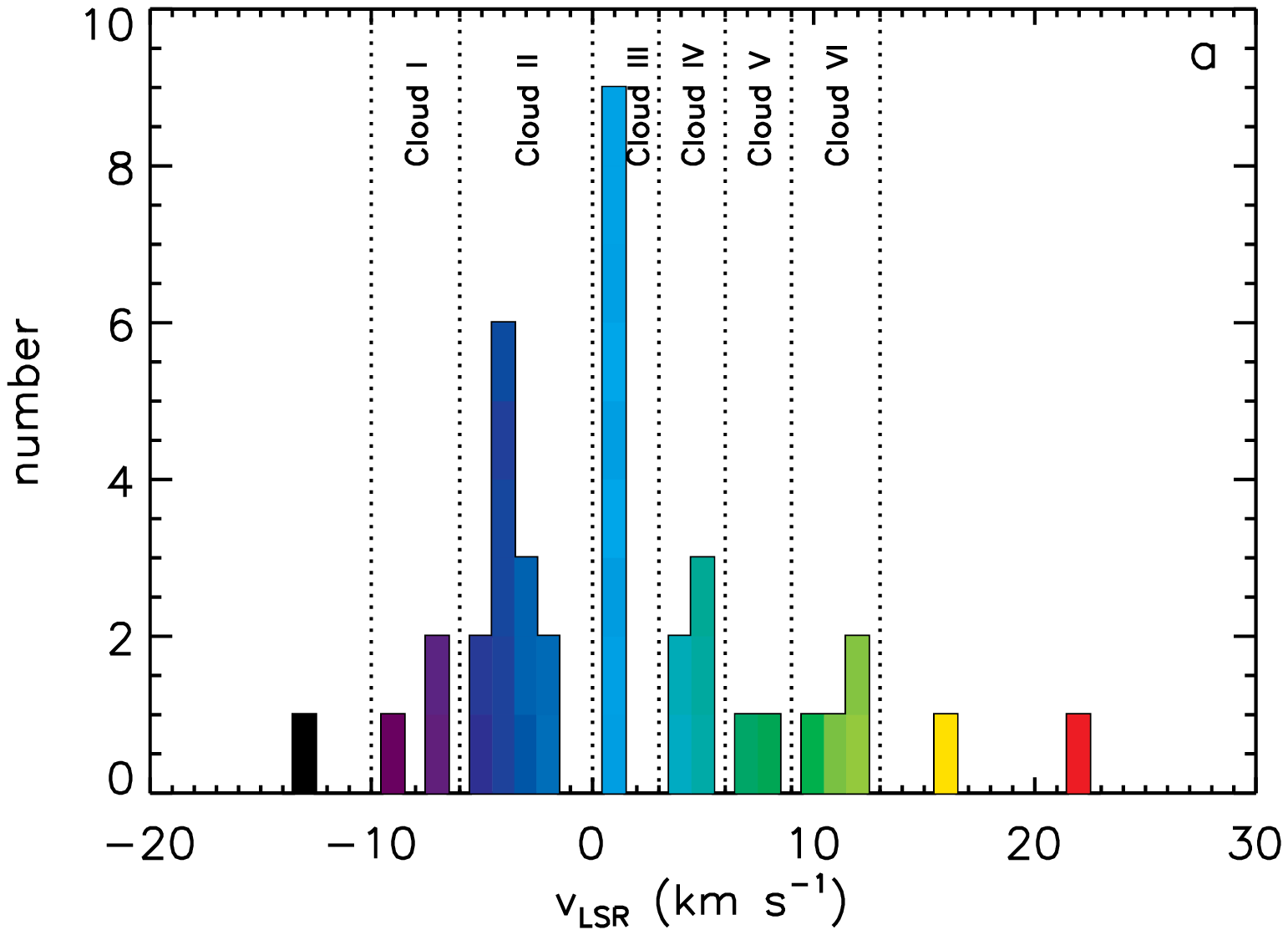}
\plotone{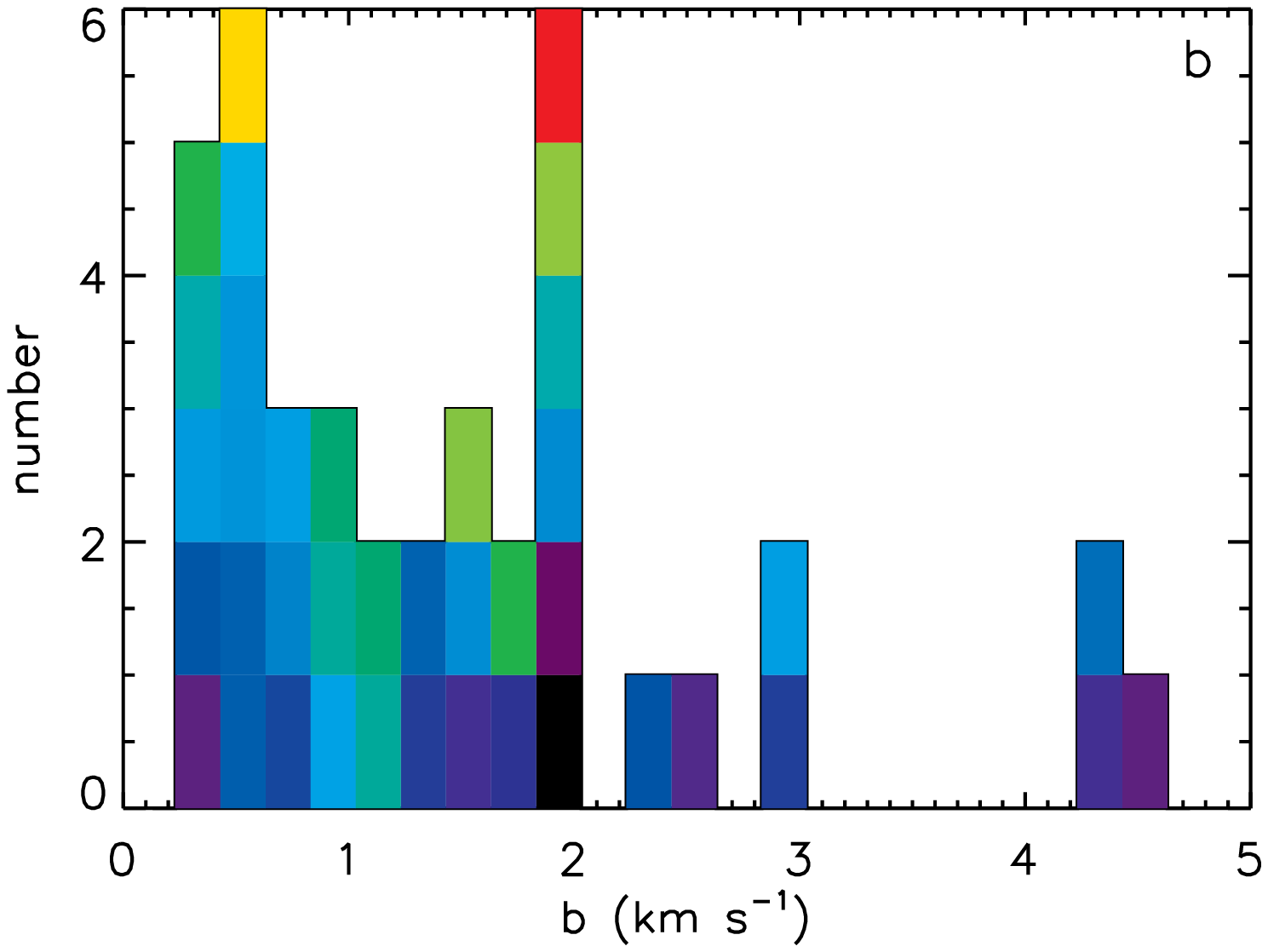}
\plotone{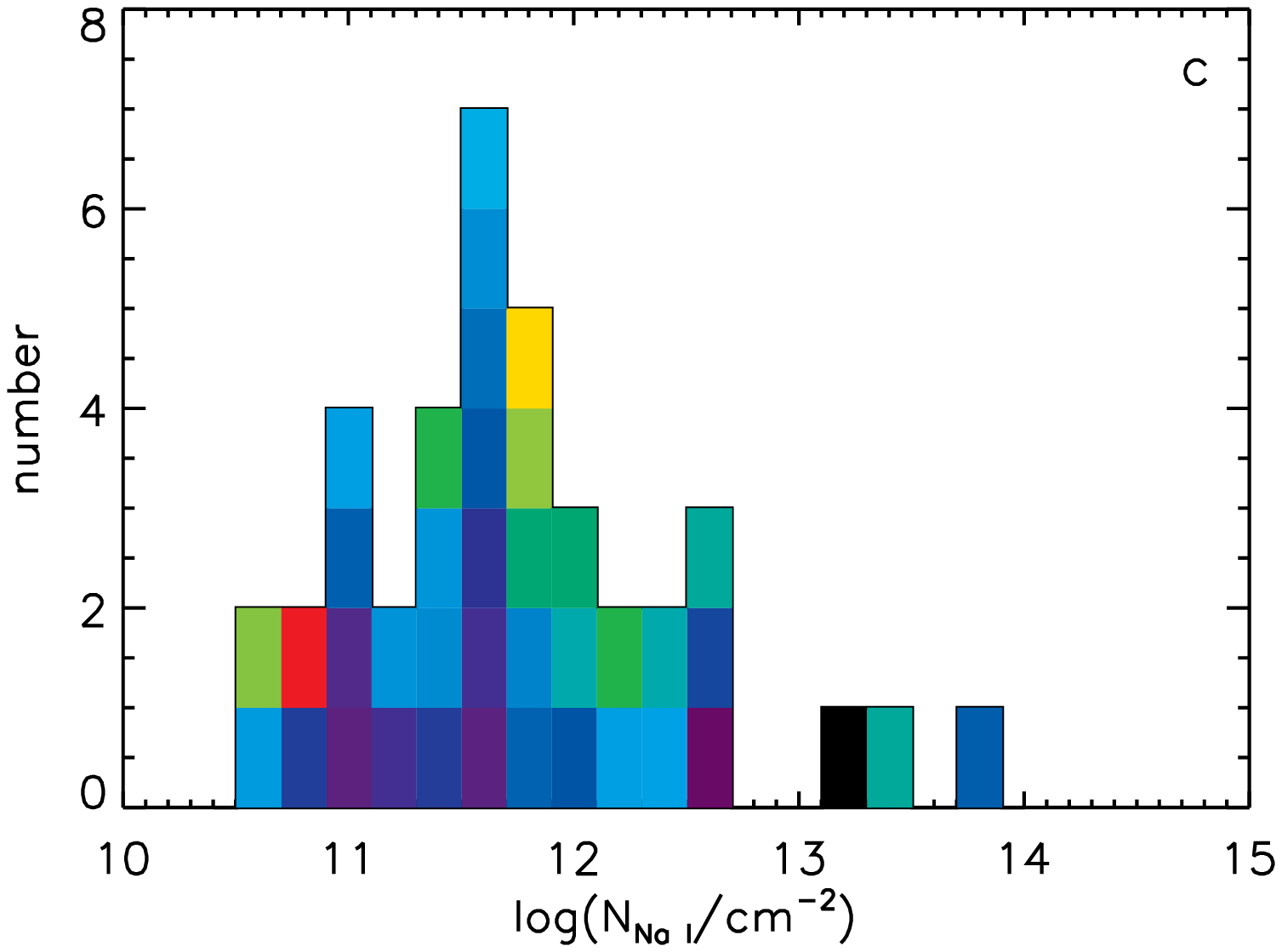}
\caption{a. Histogram showing the distribution of velocities of our detected Na~\textsc{i} components. The velocity limits of our identified Clouds I through VI are also shown (see text for more details). b. Histogram showing the distribution of Doppler parameters of our detected Na~\textsc{i} components. c. Histogram showing the distribution of column densities of our detected Na~\textsc{i} components.  Each component is colored according to its velocity in all three histograms.\label{histograms}}
\end{figure}

The fit properties for all components are given in Table \ref{naitable} for Na~\textsc{i} and \ref{kitable} for K~\textsc{i}. The distributions of the velocities, Doppler parameters, and column densities for the Na~\textsc{i} components are shown in Fig.~\ref{histograms}a, b, and c, respectively. The velocities fall into a number of clumps, which we use to separate the detected Na~\textsc{i} components into different clouds (see \S\ref{identification}). The majority of the detected ISM components have relatively narrow lines ($b<2$ km s$^{-1}$), although three components have $b\sim4.5$ km s$^{-1}$; the distribution shows two peaks, at $\sim0.5$ km s$^{-1}$ and $\sim2$ km s$^{-1}$. The distribution of Na~\textsc{i} column densities peaks at $\log(N_{\mathrm{Na~\textsc{i}}}/\mathrm{cm}^{-2})\sim11.5$, with a high column density tail of the distribution reaching to $\log(N_{\mathrm{Na~\textsc{i}}}/\mathrm{cm}^{-2})\sim14$. There does not appear to be an obvious correlation between component velocity and either Doppler parameter or column density. For comparison, the distributions of $b$ and column density that we find for Na~\textsc{i} are similar to those found by \cite{WymanRedfield} in the direction of the past solar trajectory; however, they measured a handful of components with higher Doppler parameters than we do (up to $b\sim8$ km s$^{-1}$), while we find some components with higher column densities than they did (their highest are ${\log(N_{\mathrm{Na~\textsc{i}}}/\mathrm{cm}^{-2})\sim12.3}$). 

\begin{figure}
\plotone{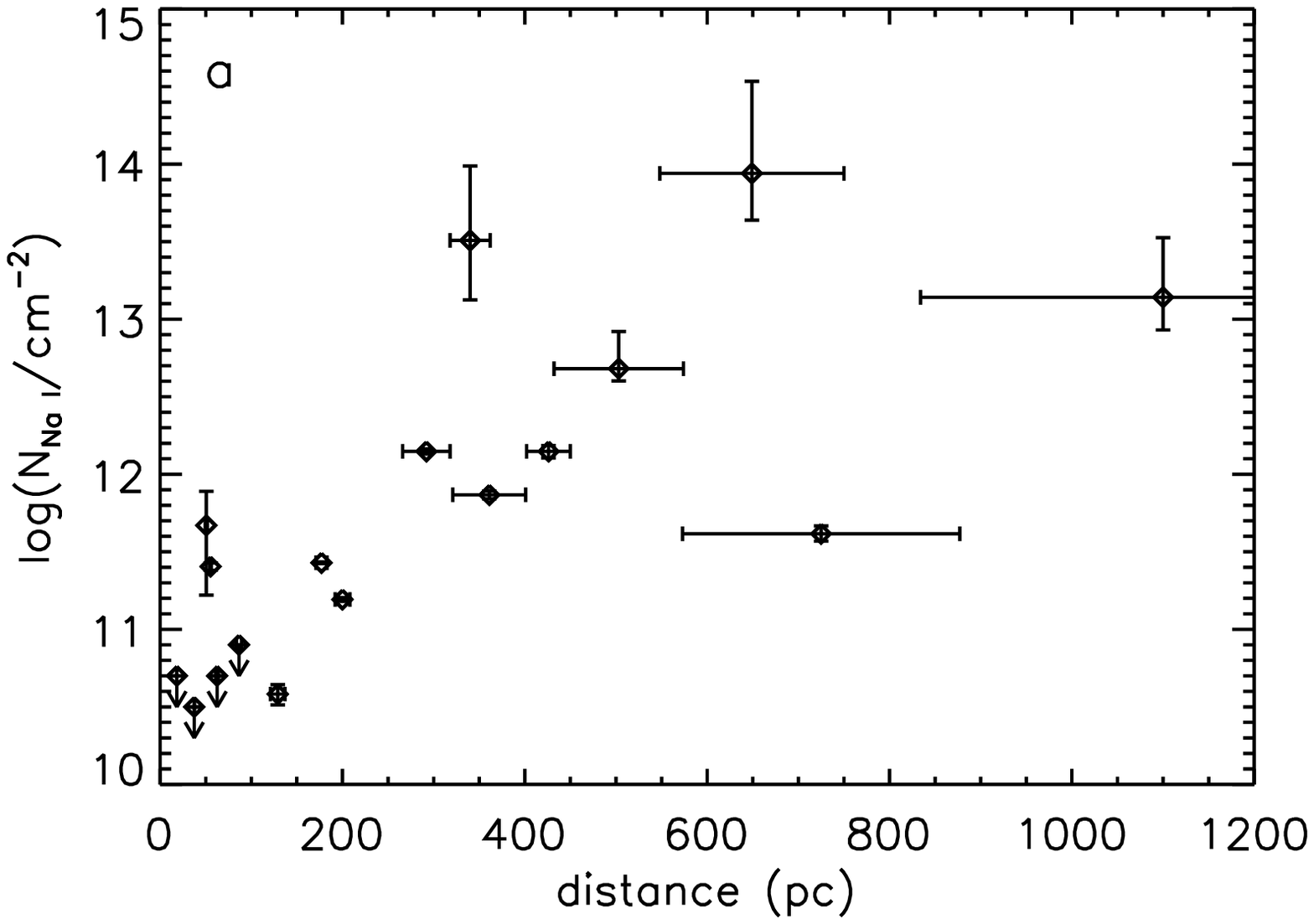}
\plotone{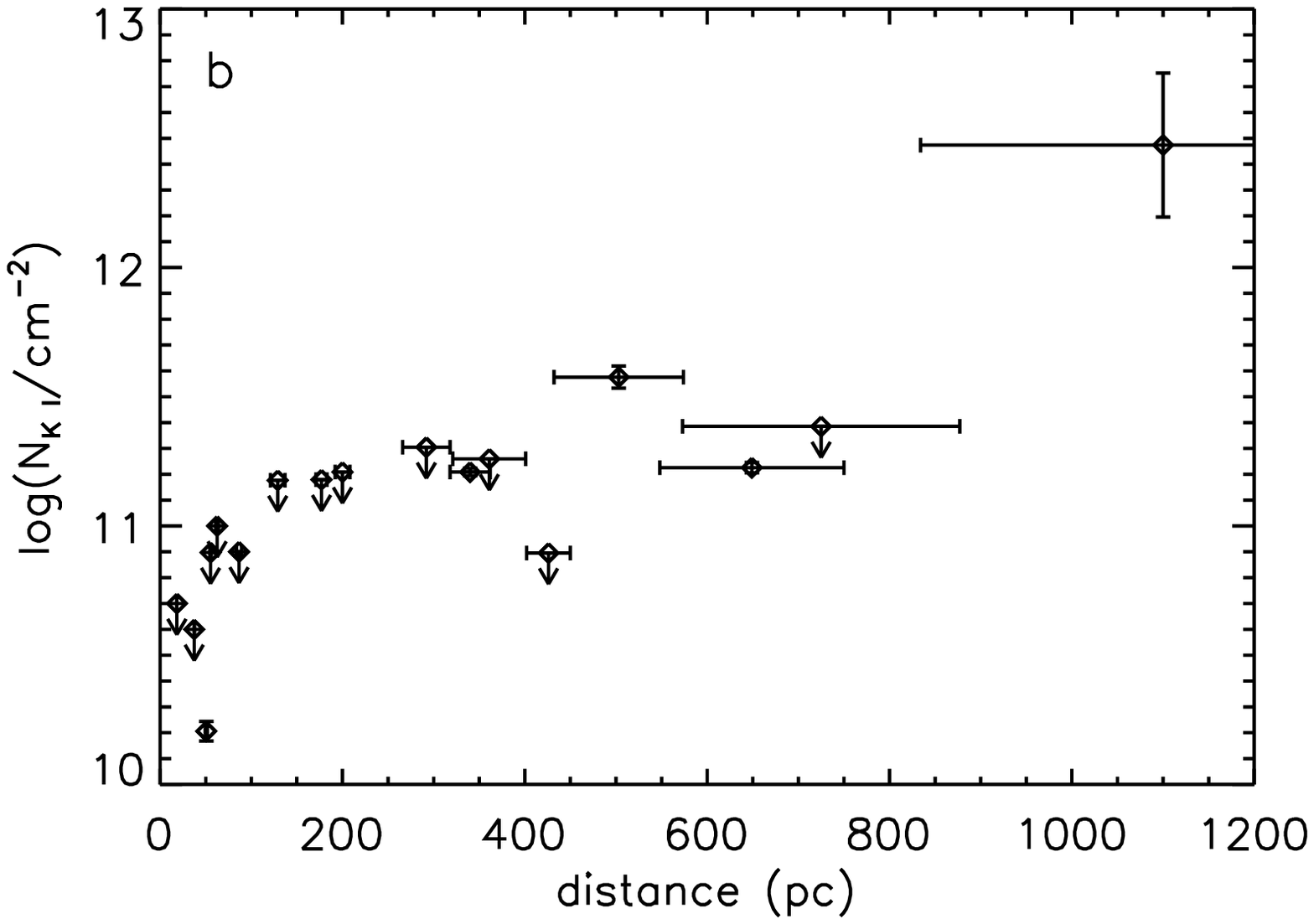}
\caption{Observed Na~\textsc{i} (a., top) and K~\textsc{i} (b., bottom) column density towards each target star as a function of distance. Some error bars are smaller than the plot points. As expected, column density increases with distance, albeit with a large amount of scatter, indicative of small-scale 
structure in the ISM. \label{na1k1column}}
\end{figure}

Fig.~\ref{na1k1column}a and b shows the total column density of Na~\textsc{i} and K~\textsc{i}, respectively, as a function of distance. As is expected, the column density of both tracers generally increases with distance. There is, however, a large amount of scatter for Na~\textsc{i}: ${\sim1}$ dex for $d<200$ pc, and greater than 2 dex at distances of ${\sim700}$ pc. This is indicative of structure in the ISM. As there are large differences in the column densities towards targets with similar distances separated by only a few degrees, there must be different distributions of Na~\textsc{i} (i.e., different structures) along these lines of sight. We have several pairs of lines of sight separated by $\sim1^{\circ}$ (corresponding to a physical separation of 1.7 pc at a distance of 100 pc or 12 pc at a distance of 700 pc), and many lines of sight showing different column densities separated by up to $14^{\circ}$ (corresponding to 25 pc at a distance of 100 pc, or 175 pc at a distance of 700 pc). Therefore, we are sensitive to ISM structure transverse to the line of sight on scales of a few to tens of parsecs.

\begin{figure*}
\plottwo{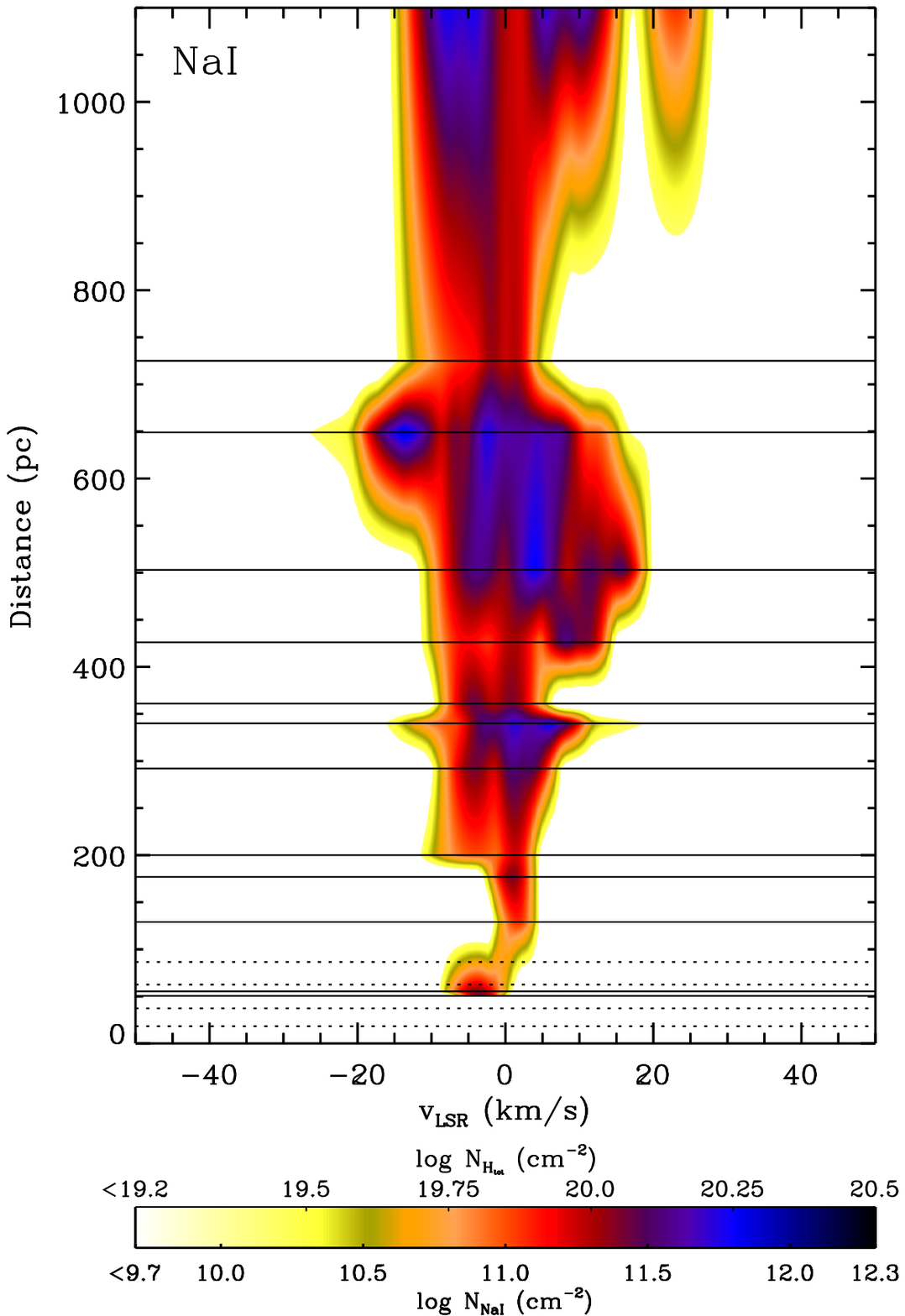}{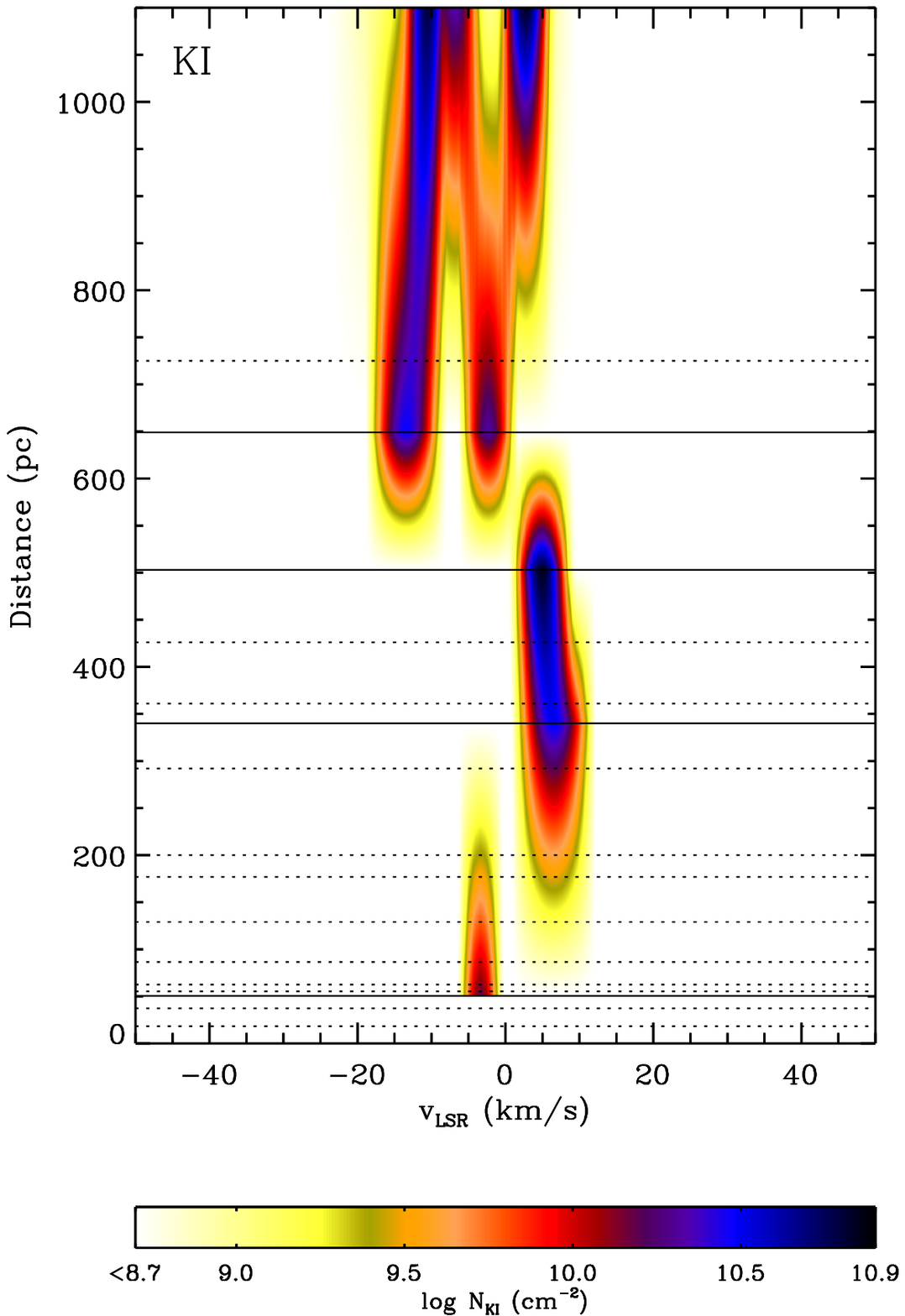}
\caption{Apparent optical depth plot for Na~\textsc{i} (left) and K~\textsc{i} (right). Each star is represented as a horizontal line; solid lines denote stars with detected Na~\textsc{i} or K~\textsc{i}, while dashed lines denote stars with no detected Na~\textsc{i} or K~\textsc{i}. The color scale denotes the column density per unit velocity in Na~\textsc{i} or K~\textsc{i}. At the location of each star this simply shows the observed absorption line profile, while elsewhere this distribution has been interpolated between the observed stars; this interpolation is necessary as we only have constraints at the locations of the horizontal lines. Note that these figures do \emph{not} show the physical distribution of ISM clouds; the length of a contour in the vertical direction is \emph{not} the physical size of the cloud along the line of sight. For example, a 1 pc thick cloud at a distance of 200 pc covering the entire {\it Kepler} field of view would manifest as a dark streak beginning at a distance of 200 pc and running unbroken to the top of the figure. Absorption shown at a given velocity and distance in this figure is caused by absorbing material that could be physically located at any smaller distance. The discontinuous colorscale at a given velocity is because of variations in the ISM across the {\it Kepler} field of view; a given star may or may not show absorption from any given cloud, depending upon its location in the plane of the sky. For Na~\textsc{i}, the column density is also converted to total H column density using the relation of \cite{WeltyHobbs01}; see also \S\ref{voldens}. \label{AOD}}
\end{figure*}

Apparent optical depth \citep{SavageSembach91} diagrams for Na~\textsc{i} and K~\textsc{i} are shown in Fig.~\ref{AOD}; see the figure caption for a full description of this plot.
These plots do indicate some continuity of cloud structure; however, due to the relatively large size of the {\it Kepler} field and the aforementioned small-scale structure of the ISM, not all components which we infer to belong to the same cloud (see \S\ref{identification}) are actually connected in this plot.

\begin{figure*}
\plotone{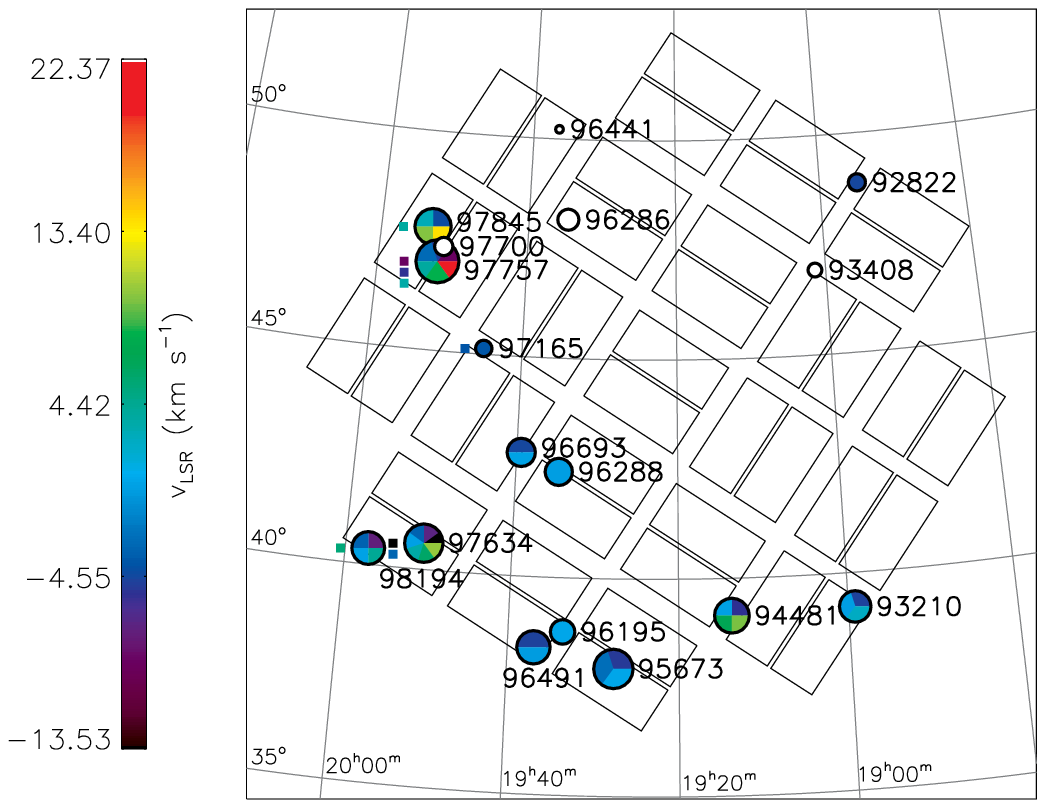}
\caption{Map of the \emph{Kepler} field, showing the observed stars and the \emph{Kepler} CCD layout. Each star is scaled according to the inverse of its distance, with larger symbols denoting more distant stars. Each plot symbol is further divided into a number of slices, each denoting one ISM component detected in Na~\textsc{i} and colored according to velocity using the scale at left, identical to that used in Fig.~\ref{histograms}; white circles denote stars with no detected Na~\textsc{i}. K~\textsc{i} detections are denoted by small squares to the left of each star, colored according to the same velocity scale. The overlayed coordinate grid shows lines of constant RA and Dec, with a spacing of 20$^\mathrm{m}$ for the RA grid and 5$^{\circ}$ for the Dec grid. The {\it Kepler} field of view is centered at $\alpha=19^h 22^m 40^s$ $\delta=+44^{\circ} 30' 00''$ \citep{KepFOV}, and each of the square CCD modules covers an area of 5 square degrees (http://kepler.nasa.gov/science/about/targetFieldOfView/). \label{Map}}
\end{figure*}

The distribution of the detected Na~\textsc{i} and K~\textsc{i} components in velocity space and on the plane of the sky in relation to the \emph{Kepler} field is shown in Fig.\ \ref{Map}. As mentioned above, most of our targets, especially those at large distances, lie on the side of the {\it Kepler} field closest to the Galactic plane (towards the lower left hand side of Fig.~\ref{Map}). Thus, we only have limited information about the ISM over the higher Galactic latitude side of the {\it Kepler} field of view. Additionally, all five of the stars for which we detect K~\textsc{i} absorption are located in the corner of the {\it Kepler} field closest to the Galactic plane.

\begin{deluxetable*}{cccccc}
\tabletypesize{\footnotesize}
\tablewidth{0pt}
\tablecaption{Na I Fit Properties \label{naitable}}
\tablehead{ 
\colhead{HIP} & \colhead{Comp.} & \colhead{$v_{\mathrm{LSR}}$ (km s$^{-1}$)} & \colhead{$b$ (km s$^{-1}$)} & \colhead{$\log(N_{\mathrm{Na~I}}/$cm$^{-2})$} & \colhead{Cloud} \\
\colhead{(1)} & \colhead{(2)} & \colhead{(3)} & \colhead{(4)} & \colhead{(5)} & \colhead{(6)} 
}

\startdata
96441 & 0 & \ldots & \ldots & $<10.7$ & \ldots \\
93408 & 0 & \ldots & \ldots & $<10.5$ & \ldots \\
97165 & 1 & $-3.120 \pm 0.010$ & $0.268 \pm 0.047$ & $11.67^{+0.22}_{-0.45}$ & II \\
92822 & 1 & $-3.96 \pm 0.32$ & $1.32 \pm 0.24$ & $11.405^{+0.023}_{-0.024}$ & II \\
97700 & 0 &\ldots&\ldots& $<10.7$ & \ldots \\
96286 & 0 &\ldots&\ldots& $<10.9$ & \ldots \\
96195 & 1 & $1.42 \pm 0.35$ & $0.231 \pm 0.052$ & $10.583^{+0.060}_{-0.070}$ & III \\
96288 & 1 & $0.89 \pm 0.11$ & $0.578 \pm 0.029$ & $11.4296 \pm 0.0061$ & III \\
96693 & 1 & $-4.04 \pm 0.53$ & $2.9 \pm 1.1$ & $10.789 \pm 0.019$ & II \\
 \ldots  & 2 & $1.24 \pm 0.13$ & $0.64 \pm 0.46$ & $10.975^{+0.015}_{-0.016}$ & III \\
93210 &  1 & $-4.19 \pm 0.20$ & $1.60 \pm 0.10$ & $11.574^{+0.047}_{-0.053}$ & II \\
 \ldots  &   2 & $0.867 \pm 0.092$ & $0.76 \pm 0.23$ & $11.754^{+0.017}_{-0.018}$ & III \\
 \ldots  &   3 & $3.65 \pm 0.16$ & $0.55 \pm 0.13$ & $11.664^{+0.023}_{-0.024}$ & IV \\
98194 & 1 & $-7.00 \pm 0.49$ & $4.46 \pm 0.32$ & $10.909^{+0.086}_{-0.107}$ & I \\
 \ldots  & 2 & $-2.538 \pm 0.058$ & $1.35 \pm 0.10$ & $11.890^{+0.020}_{-0.021}$ & II \\
 \ldots  & 3 & $1.218 \pm 0.024$ & $0.99\pm 0.13$ & $12.496^{+0.073}_{-0.088}$ & III \\
\ldots & 4 & $5.369 \pm 0.051$ & $1.11 \pm 0.11$ & $13.45^{+0.52}_{-0.48}$ & IV \\
96491 & 1 & $-4.16 \pm 0.50$ & $1.65 \pm 0.22$ & $11.621^{+0.041}_{-0.045}$ & II \\
\ldots   & 2 & $0.81 \pm 0.19$ & $1.47 \pm 0.19$ & $11.504 \pm 0.015$ & III \\

94481 &   1 & $-4.94 \pm 0.27$ & $2.45 \pm 0.77$ & $11.10^{+0.27}_{-0.92}$ & II \\
\ldots   &   2 & $1.079 \pm 0.022$ & $1.92 \pm 0.37$ & $11.327^{+0.067}_{-0.080}$ & III \\
 \ldots  &   3 & $8.194 \pm 0.087$ & $0.94 \pm 0.12$ & $11.902 \pm 0.014$ & V \\
\ldots   &  4 & $11.281 \pm 0.038$ & $0.298 \pm 0.067$ & $11.429^{+0.080}_{-0.098}$ & VI \\
97845 & 1 & $-3.74 \pm 0.34$ & $2.38 \pm 0.23$ & $12.060^{+0.071}_{-0.085}$ & II \\
  \ldots & 2 & $3.98 \pm 0.15$ & $1.86 \pm 0.54$ & $12.39^{+0.38}_{-0.15}$ & IV \\
 \ldots  & 3 & $11.588 \pm 0.095$ & $1.90 \pm 0.17$ & $11.7897 \pm 0.0017$ & VI \\
 \ldots  & 4 & $15.75 \pm 0.14$ & $0.44 \pm 0.23$ & $11.76^{+0.39}_{-0.29}$ & \ldots \\
97634 & 1 & $-13.53 \pm 0.11$ & $1.882 \pm 0.072$ & $13.140^{+0.089}_{-0.113}$ & \ldots \\
 \ldots  & 2 & $-6.639 \pm 0.065$ & $0.249 \pm 0.099$ & $11.56^{+0.16}_{-0.26}$ & I \\
 \ldots  & 3 & $-2.59 \pm 0.15$ & $0.50 \pm 0.33$ & $13.84^{+0.67}_{-0.43}$ & II \\ 
\ldots   & 4 & $1.09 \pm 0.32$ & $2.94 \pm 0.33$ & $12.218 \pm 0.056$ & III \\
  \ldots & 5 & $4.681 \pm 0.012$ & $0.378 \pm 0.072$ & $12.039^{+0.066}_{-0.078}$ & IV \\
 \ldots  & 6 & $7.348 \pm 0.018$ & $1.119 \pm 0.025$ & $11.997 \pm 0.016$ & V \\
 \ldots  & 7 & $11.91 \pm 0.39$ & $1.47 \pm 0.41$ & $10.503 \pm 0.037$ & VI \\
95673 &  1 & $-4.645 \pm 0.056$ & $4.264 \pm 0.052$ & $11.186^{+0.030}_{-0.032}$ & \ldots \\
  \ldots &  2 & $-1.73 \pm 0.21$ & $0.44 \pm 0.11$ & $11.10^{+0.14}_{-0.16}$ & II \\
 \ldots  &  3 & $1.32 \pm 0.38$ & $0.56 \pm 0.37$ & $11.126^{+0.042}_{-0.046}$ & III \\ 
97757 & 1 & $-8.63 \pm 0.13$ & $1.92 \pm 0.22$ & $12.53 \pm 0.44$  & I \\
 \ldots  & 2 & $-4.39 \pm 0.49$ & $0.83 \pm 0.25$ & $12.55 \pm 0.79$ & II \\
\ldots   & 3 & $-2.13 \pm 0.95$ & $4.4 \pm 3.8$ & $11.675^{+0.091}_{-0.115}$ & \ldots \\
\ldots   & 4 & $4.939 \pm 0.066$ & $0.866 \pm 0.096$ & $12.67^{+0.26}_{-0.75}$ & IV \\
\ldots   & 5 & $9.63 \pm 0.12$ & $1.71 \pm 0.13$ & $12.228^{+0.013}_{-0.014}$ & VI \\
 \ldots  & 6 & $22.374 \pm 0.095$ & $1.87 \pm 0.73$ & $10.786^{+0.073}_{-0.088}$ & \ldots 

\enddata

\tablecomments{(1) HIP catalog designation, in order of increasing distance. (2) Number of detected Na~\textsc{i} component for that star (numbering from most negative to most positive velocity), zero if no Na~\textsc{i} is detected towards that star. (3) LSR velocity of component. (4) Doppler parameter of component. (5) Logarithmic Na~\textsc{i} column density of component. (6) Cloud assignment of component (see text for details).}

\end{deluxetable*}

\begin{deluxetable*}{ccccccc}
\tabletypesize{\footnotesize}
\tablewidth{0pt}
\tablecaption{K I Fit Properties \label{kitable}}
\tablehead{ 
\colhead{HIP} &  \colhead{Comp.} & \colhead{$v_{\mathrm{LSR}}$ (km s$^{-1}$)} & \colhead{$b$ (km s$^{-1}$)} & \colhead{$\log(N_{\mathrm{K~I}}/$cm$^{-2})$} & \colhead{$\Delta v$ (km s$^{-1}$)} & \colhead{Cloud} \\
\colhead{(1)} & \colhead{(2)} & \colhead{(3)} & \colhead{(4)} & \colhead{(5)} & \colhead{(6)} & \colhead{(7)} 
}

\startdata
96441 &  0 &\ldots&\ldots& $<10.7$ &\ldots&\ldots\\
93408 &  0 &\ldots&\ldots& $<10.6$ &\ldots&\ldots\\
97165 &  1 & $-3.214 \pm 0.051$ & $0.1008 \pm 0.0035$ & $10.206 \pm 0.038$ & $0.014 \pm 0.052$ & II \\
92822 &  0 &\ldots&\ldots& $<10.8$ &\ldots&\ldots\\
97700 &  0 &\ldots&\ldots& $<11.0$ &\ldots&\ldots\\
96286 &  0 &\ldots&\ldots& $<10.9$ &\ldots&\ldots\\
96195 &  0 &\ldots&\ldots& $<11.0$ &\ldots&\ldots\\
96288 &  0 &\ldots&\ldots& $<11.1$ &\ldots&\ldots\\
96693 &  0 &\ldots&\ldots& $<11.1$ &\ldots&\ldots\\
93210 &  0 &\ldots&\ldots& $<11.0$ &\ldots&\ldots\\
98194 &  1 & $6.54 \pm 0.11$ & $1.66 \pm 0.19$ & $10.847 \pm 0.032$  & $-1.17 \pm 0.12$ & IV \\
96491 &  0 &\ldots&\ldots& $<11.0$ &\ldots&\ldots\\
94481 &  0 &\ldots&\ldots& $<10.6$ &\ldots&\ldots\\
97845 &  1 & $4.96 \pm 0.069$ & $0.66 \pm 0.15$ & $11.377 \pm 0.066$ & $-0.97 \pm 0.17$ & IV \\
97634 &  1 & $-13.43 \pm 0.16$ & $1.54 \pm 0.29$ & $10.788 \pm 0.042$ & $0.10 \pm 0.19$ & \ldots \\
   \ldots    & 2 & $-2.31 \pm 0.31$ & $0.94 \pm 0.60$ & $10.394 \pm 0.078$ & $-0.28 \pm 0.34$ & II \\
95673 &  0 &\ldots&\ldots& $<11.0$ &\ldots&\ldots\\
97757 &  1 & $-8.743 \pm 0.060$ & $0.202 \pm 0.038$ & $12.39 \pm 0.32$ & $0.12 \pm 0.14$ & I \\
\ldots& 2 & $-4.90 \pm 0.23$ & $0.88 \pm 0.51$ & $10.420 \pm 0.087$ & $0.51 \pm 0.54$ & II \\
\ldots& 3 & $4.633 \pm 0.024$ & $0.393 \pm 0.075$ & $11.59 \pm 0.11$ & $0.306 \pm 0.070$ & IV

\enddata

\tablecomments{(1) HIP catalog designation, in order of increasing distance. (2) Number of detected K~\textsc{i} component for that star (numbering from most negative to most positive velocity), zero if no K~\textsc{i} is detected towards that star. Note that this is not the same as the component number of the corresponding Na~\textsc{i} component listed in Table \ref{naitable}. (3) LSR velocity of component. (4) Doppler parameter of component. (5) Logarithmic K~\textsc{i} column density of component. (6) Velocity difference between associated Na~\textsc{i} and K~\textsc{i} components, i.e., $\Delta v=v_{\mathrm{Na~\textsc{i}}}-v_{\mathrm{K~\textsc{i}}}$. (7) Cloud assignment of component (see text for details).}

\end{deluxetable*}

\subsection{Identification of ISM Clouds}
\label{identification}

Due to our relatively sparse spatial coverage we concentrate on identifying ISM clouds based on the line velocity centroids. We assume that clouds move as approximately solid bodies, 
 such that if we see two stars with absorption at the same LSR velocity at opposite sides of the field of view, they would result from the same cloud. In principle there could always be more distant absorption components which happen to share the same velocity as an actual cloud, and would thus be misidentified as belonging to the nearer cloud. Given our sparse spatial coverage we cannot positively identify such contamination, so we will proceed with the above assumptions with the caveat that such velocity interlopers may be a problem on longer sightlines. An additional complication is caused by Galactic rotation. At a distance of 426 pc (beyond which we only have four targets), objects at rest with respect to Galactic rotation towards our highest and lowest Galactic latitude targets will show radial velocities differing by $\sim2.5$ km s$^{-1}$ \citep[using Eqn. 5.170 of][]{Lang99}. This velocity differential is large enough to potentially cause confusion regarding the cloud identifications. Most components, however, would show velocity shifts due to Galactic rotation of $<2.5$ km s$^{-1}$ due to smaller distances and less widely separated sightlines, and so we again proceed with the caveat that radial velocity differentials due to Galactic rotation could be an additional source of confusion for our cloud identifications. In order to set distance limits on the clouds, we define the lack of absorption at a given velocity as the lack of presence of a cloud. There could still be absorption that is too weak for us to detect; typically, we are sensitive to column densities of $\log(N/\mathrm{cm}^{-2})\gtrsim10.7$ for Na~\textsc{i} and $\log(N/\mathrm{cm}^{-2})\gtrsim11.0$ for K~\textsc{i}. Lack of detection of absorption could be due to either low densities, resulting in a column density of Na or K below our detection limits, or the presence of Na or K but in higher ionization states (Na~\textsc{ii}, K~\textsc{ii}, etc.), indicating high temperatures or the presence of ionizing radiation. Thus, we are only sensitive to a certain type of cloud: relatively cold ($T<1000$ K), predominantly neutral clouds \citep[e.g.,][]{Hobbs78,Lallement}. Conveniently, this is the type of cloud that is capable of significantly compressing astrospheres. Previous works to map the ISM using absorption line measurements have made similar assumptions \citep[e.g.,][]{Lallement}.

\begin{figure*}
\plotone{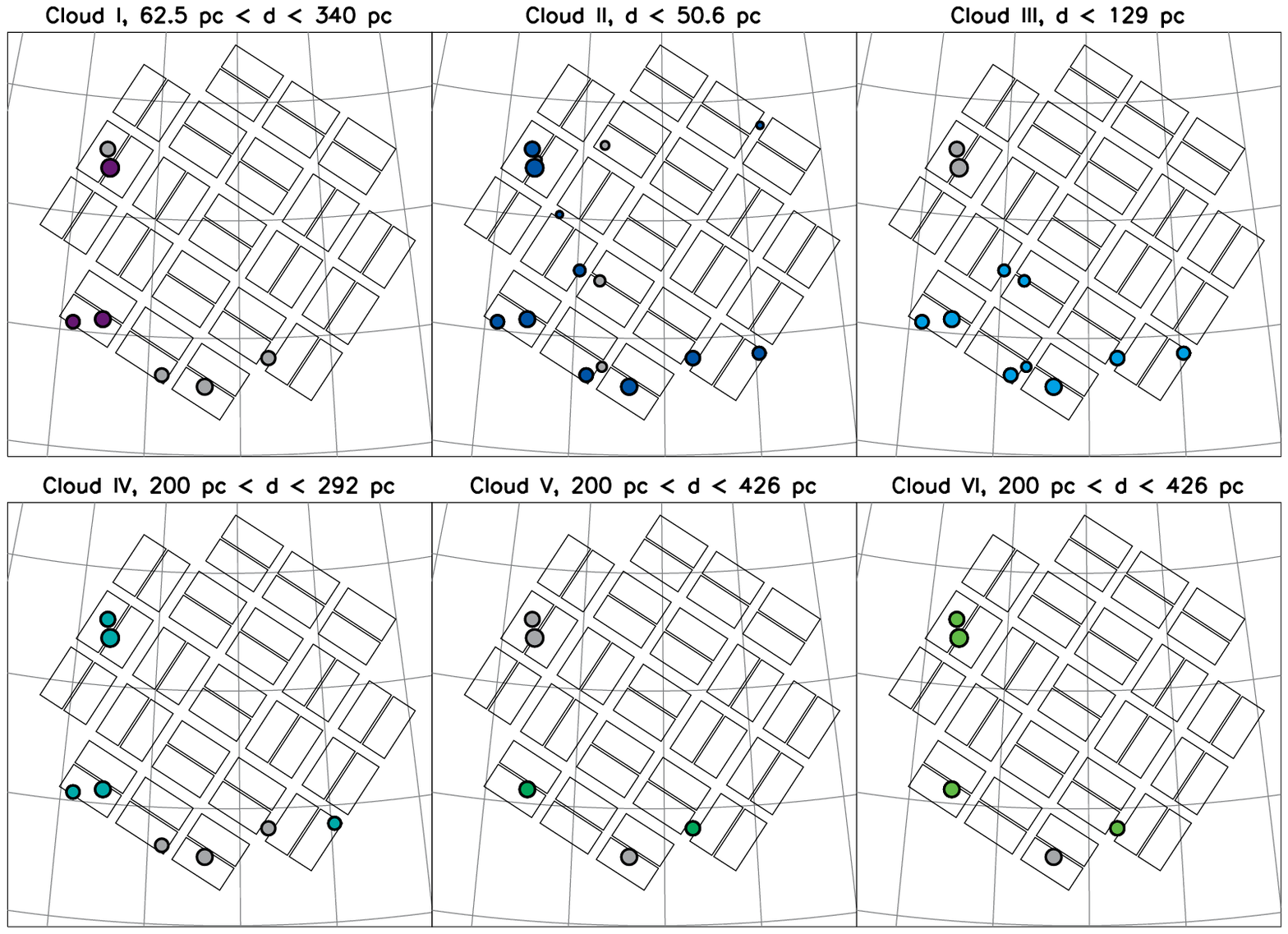}
\caption{Same as Fig.\ \ref{Map}, but showing each cloud identified in \S \ref{identification}. For each cloud only the stars more distant than the closest detection of that cloud are shown. Stars with absorption detected from that cloud are colored by velocity according to the same scale as in Fig.~\ref{Map}, while those with no absorption detected are shown in gray, in order to show the spatial continuity of each cloud. 
 The coordinate grid is the same as in Fig.~\ref{Map}. \label{CloudMap}}
\end{figure*}

As is apparent in Fig.\ \ref{histograms}a, there are six distinct clumps in the distribution: between $-10$ and $-6$, $-5$ and $-1$, $0$ and $+2$, $+3$ and $+6$, $+6$ and $+9$, and $+9$ and $+13$ km s$^{-1}$. Each of these we identify as clouds due to their coincidence and isolation in the velocity space, and their spacial continuity (see Figs.~\ref{Map} and \ref{CloudMap}). 
We therefore identify eleven ISM clouds in the \emph{Kepler} field of view, five of which are only seen in absorption in the spectrum of one star. We attribute the other thirty-four detected Na~\textsc{i} components to six other clouds, within the velocity ranges defined above.

We will now discuss the various identified clouds and single components, in order from most negative to most positive velocities. The bluest component is seen in the spectrum of HIP 97634 and is detected in both Na~\textsc{i} and K~\textsc{i}, with LSR velocities of $-13.53$ km s$^{-1}$ (Na~\textsc{i}) and $-13.43$ km s$^{-1}$ (K~\textsc{i}). As there are no components detected near this velocity for other stars, we can only set an upper limit on the distance to this cloud of 649 pc, the distance to HIP 97634. The spatial extent is largely unconstrained, but it is not seen in the spectra of the only two more distant stars, HIP 95673 and 97757, limiting its extent towards the northeast and southeast sides of the \emph{Kepler} field of view (the upper and lower left portion of Fig.~\ref{Map}, respectively).

Next are the three Na~\textsc{i} components between $-10$ and $-6$ km s$^{-1}$ (seen in the spectra of HIP 97634, 97757, and 98194), all located near the eastern corner of the field of view. Due to the spatial coincidence of these components, we identify this as ``Cloud I.'' The upper left panel of Fig. \ref{CloudMap} shows the spatial distribution of these stars. We do not have any targets located in between these three stars, but no absorption in this velocity range is seen in the spectra of HIP 97700, located only 22' from HIP 97757. We thus consider the distance to HIP 97700, 62.5 pc, to be the lower limit on the distance to this cloud, while the distance to the closest star with detected absorption, HIP 98194 (340 pc), is the upper limit.
Given the maximum separation between stars showing absorption from Cloud I ($6.7^{\circ}$), we can estimate the size of the cloud in the plane of the sky as 7.3 pc and 40 pc if located at the minimum and maximum distances, respectively. We note that this is a lower limit on the size, as the cloud may extend beyond the region probed by our survey. One Cloud I component, that towards HIP 97757, is also detected in K~\textsc{i}.

Next is the largest clump in velocity space, eleven stars with absorption between $-5$ and $-1$ km s$^{-1}$ (HIP 92822, 93210, 94481, 95673, 96491, 96693, 97165, 97634, 97757, 97845, and 98194), distributed across the field of view. There exist many stars with no absorption detected, with locations on the plane of the sky near or in between less distant stars with detected absorption. For instance, Na~\textsc{i} absorption is not detected in the spectrum of HIP 97700 or 96288, at distances of 62.5 and 177 pc, respectively, but is detected in the spectrum of HIP 97165, located almost directly between HIP 97700 and 96288 in the plane of the sky, at a distance of 50.6 pc.

We suggest two possibilities to explain the above observations. First, this absorption may be due to one cloud, which contains a large amount of substructure. Alternatively, this absorption may be due to two or more smaller clouds which happen to share similar velocities. 
 We do not have sufficient spatial or depth resolution to distinguish between these possibilities, so for the time being we designate all the absorption between $-5$ and $-1$ km s$^{-1}$ as ``Cloud II,'' with the caveat that this may not be one monolithic cloud. Three Cloud II components (those towards HIP 97165, HIP 97634, and HIP 97757) are also detected in K~\textsc{i}.

Due to the aforementioned substructure we cannot set a robust lower limit on the distance to this cloud, but the closest star with detected absorption is HIP 97165, at a distance of 50.6 pc. Thus, at least a part of this cloud is located at $d<50.6$ pc, and is therefore the closest ISM structure found in this work and is located within the Local Bubble \citep[the wall of the Local Bubble is located at a distance of $\sim100-150$ pc in this direction;][]{Lallement}. The detection of this cloud at a distance of $\sim50$ pc by \cite{Lallement} is discussed in more detail in \S\ref{comparison}. Given the distance limits to the cloud and the separation of $13.9^{\circ}$ between the most widely separated stars showing Cloud II absorption, we can estimate a size of the cloud of 12.5 pc if located at the maximum distance of 50.6 pc (smaller if located closer).

We also detect very broad, shallow components in the spectra of HIP 97757 and HIP 95673 at velocities appropriate for Cloud II. 
 In the case of HIP 97757 this component is not well separated from the other, stronger components seen in the spectrum, but its inclusion in the fit is supported by the F-test. These components have two of the largest Doppler parameters of any of our detected components ($b=4.4 \pm 3.8$ km s$^{-1}$ for HIP 97757 and $4.264 \pm 0.052$ for HIP 95673). While these components could be produced by a single very hot or turbulent cloud, the very large Doppler parameters make this seem unlikely; these components may instead be caused by a number of narrower, unresolved absorption features, unsurprising as HIP 95673 and HIP 97757 are our two most distant targets, at 725 and 1100 pc, respectively.
 For these reasons we attribute the stronger, narrower Na~\textsc{i} components in the spectra of these stars in the Cloud II velocity range to Cloud II, rather than these broad components. 

As is apparent from Fig.\ \ref{histograms}a, another, very narrow clump in the velocity distribution exists between $0$ and $+2$ km s$^{-1}$; the bluest and reddest components in this group are separated by only 0.6 km s$^{-1}$ in velocity space. Na~\textsc{i} absorption at these velocities is seen in the spectra of HIP 93210, 94481, 95673, 96195, 96288, 96491, 96693, 97634, and 98194.  The spatial distribution of these stars is presented in the upper right panel of Fig.\ \ref{CloudMap}. Again like Cloud I these stars form a spatially coherent group. We thus designate this as ``Cloud III.''  All stars within the southeastern part of the \emph{Kepler} field of view show absorption at these velocities, and so we can only set an upper limit on the distance to this cloud of 129 pc, the distance to the nearest star with detected absorption, HIP 96195. The maximum separation between stars with Cloud III detections is $11.2^{\circ}$, and so we can estimate a size limit of 25.3 pc if the cloud is located at the maximum distance of 129 pc. None of these components are detected in K~\textsc{i}.

Absorption between $+3$ and $+6$ km s$^{-1}$ is detected in the spectra of five stars (HIP 93210, 97634, 97757, 97845, and 98194). The spatial extent of these stars is shown in the lower left panel of Fig.\ \ref{CloudMap}. We refer to this as ``Cloud IV.'' Based upon the distances to the closest detection (HIP 93210) and the farthest non-detection in the region delineated by the detections (HIP 96693), we can set distance limits of 200 pc $< d <$ 292 pc for Cloud IV. With a maximum separation of $13.0^{\circ}$ between stars with Cloud IV detections, we can estimate sizes of 46.2 pc and 67.4 pc if located at the minimum and maximum distances, respectively. Three of the five Cloud IV components, those towards HIP 97757, 97845, and 98194, are also detected in K~\textsc{i}.

Two stars (HIP 94481 and 97634) show Na~\textsc{i} absorption between $+7$ and $+9$ km s$^{-1}$. As can be seen in the lower middle panel of Fig.\ \ref{CloudMap}, these two stars both lie near the southern edge (bottom part of the figure) of the \emph{Kepler} field of view, and there are no more distant stars with no absorption at these velocities located between these two stars. Thus, we tentatively label this as ``Cloud V,'' despite the detection towards only two stars; we note that in principle this could be two separate clouds which share the same velocity, but we will proceed under the assumption that this is a single cloud. The maximum distance to Cloud V is 426 pc, the distance to HIP 94481. We do not have any targets located directly between HIP 94481 and 97634, and so we cannot directly set a lower limit on the distance to this cloud. However, the cloud is unlikely to cover only the region of the sky directly in between the two stars. It cannot extend too far to the south, as no absorption is seen in the spectrum of the more distant HIP 95673 at these velocities. If we assume instead that HIP 94481 and 97634 are located near the southern edge of Cloud V, we can set a tentative lower limit of 200 pc through the non-detection of the cloud towards HIP 96288 and 96693. The separation of 7.2$^{\circ}$ between HIP 94481 and 97634 corresponds to cloud sizes of 25.3 pc at a distance of 200 pc and 53.8 pc at a distance of 426 pc.

Four stars (HIP 94481, 97634, 97757, and 97845) display Na~\textsc{i} absorption between $+9$ and $+12$ km s$^{-1}$. We attribute these four components to one cloud, ``Cloud VI.'' The spatial distribution of these stars is shown in the bottom right panel of Fig.\ \ref{CloudMap}. Based upon the closest detection and the nearest non-detection within this region, we set distance limits of 200 pc $< d <$ 426 pc. The maximum separation between stars with Cloud VI detections is $8.77^{\circ}$, and so we can set limits on the transverse size of 30.7 pc and 65.3 pc if located at the minimum and maximum distances, respectively. No K~\textsc{i} is detected for any of these components.

Another single component is seen in the spectrum of HIP 97845 at a velocity of $15.75$ km s$^{-1}$. This cloud would lie at a maximum distance of 503 pc. Finally, the reddest component is detected in absorption towards HIP 97757, at a velocity of $22.374$ km s$^{-1}$ and a maximum distance of 1100 pc.

The ranges of parameters seen in each of our identified clouds with more than one observed sightline are listed in Table \ref{meantable}. 

\subsection{Comparison of Na~\textsc{i} and K~\textsc{i}}

All eight detected K~\textsc{i} components lie within 1.2 km s$^{-1}$ of a detected Na~\textsc{i} component. Five of the eight have velocities identical to the corresponding Na~\textsc{i} component to within $1\sigma$. Almost without exception the Na~\textsc{i} components with a corresponding K~\textsc{i} component are those with the highest column densities for a given star. Three K~\textsc{i} components are associated with Cloud IV, three with Cloud II, one with Cloud I, and one with a cloud detected along only one sightline. Additionally, the sightlines with detected K~\textsc{i} absorption tend to be those located closer to the Galactic plane (off the lower left side of Fig.\ \ref{Map}).

\begin{figure}
\plotone{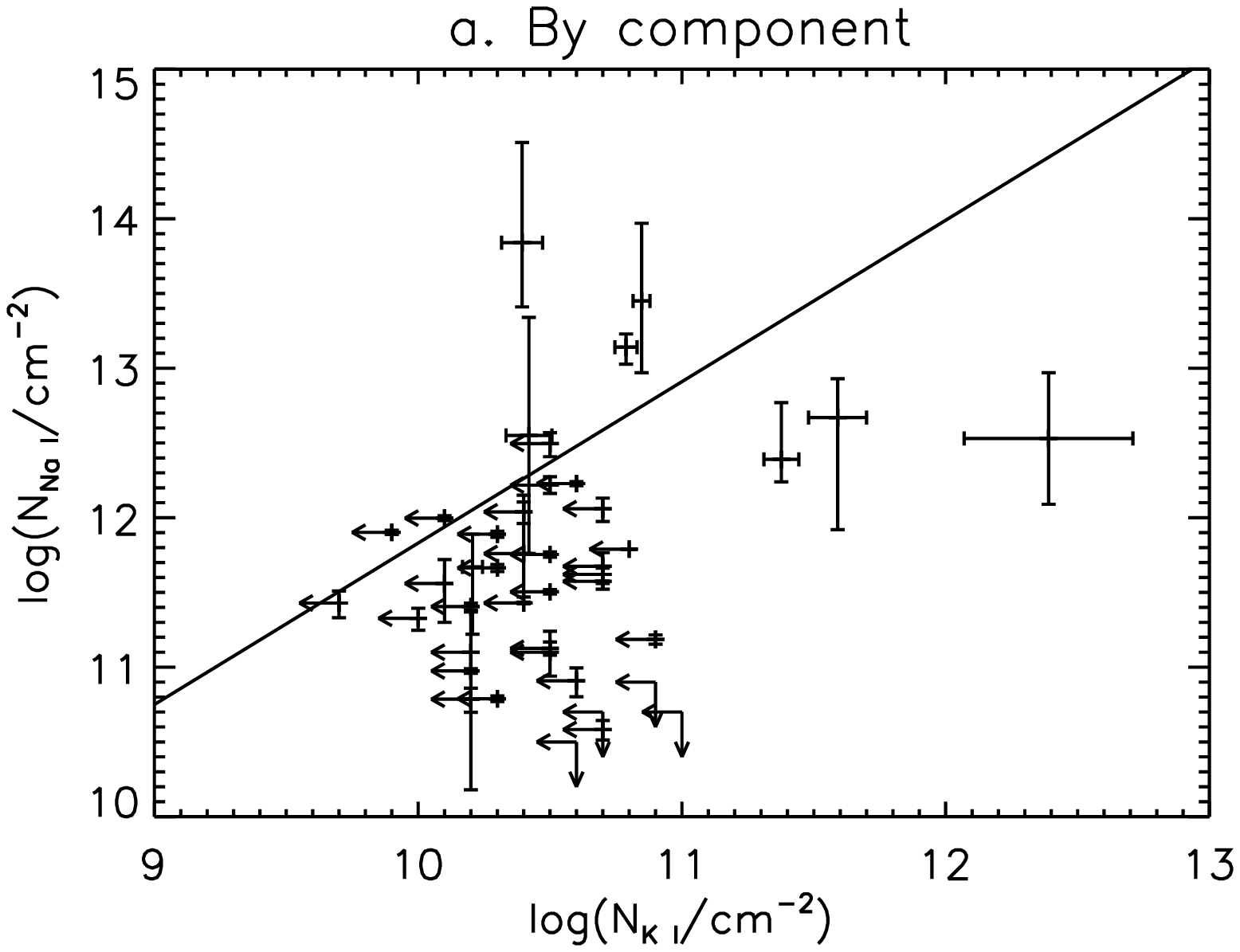}
\plotone{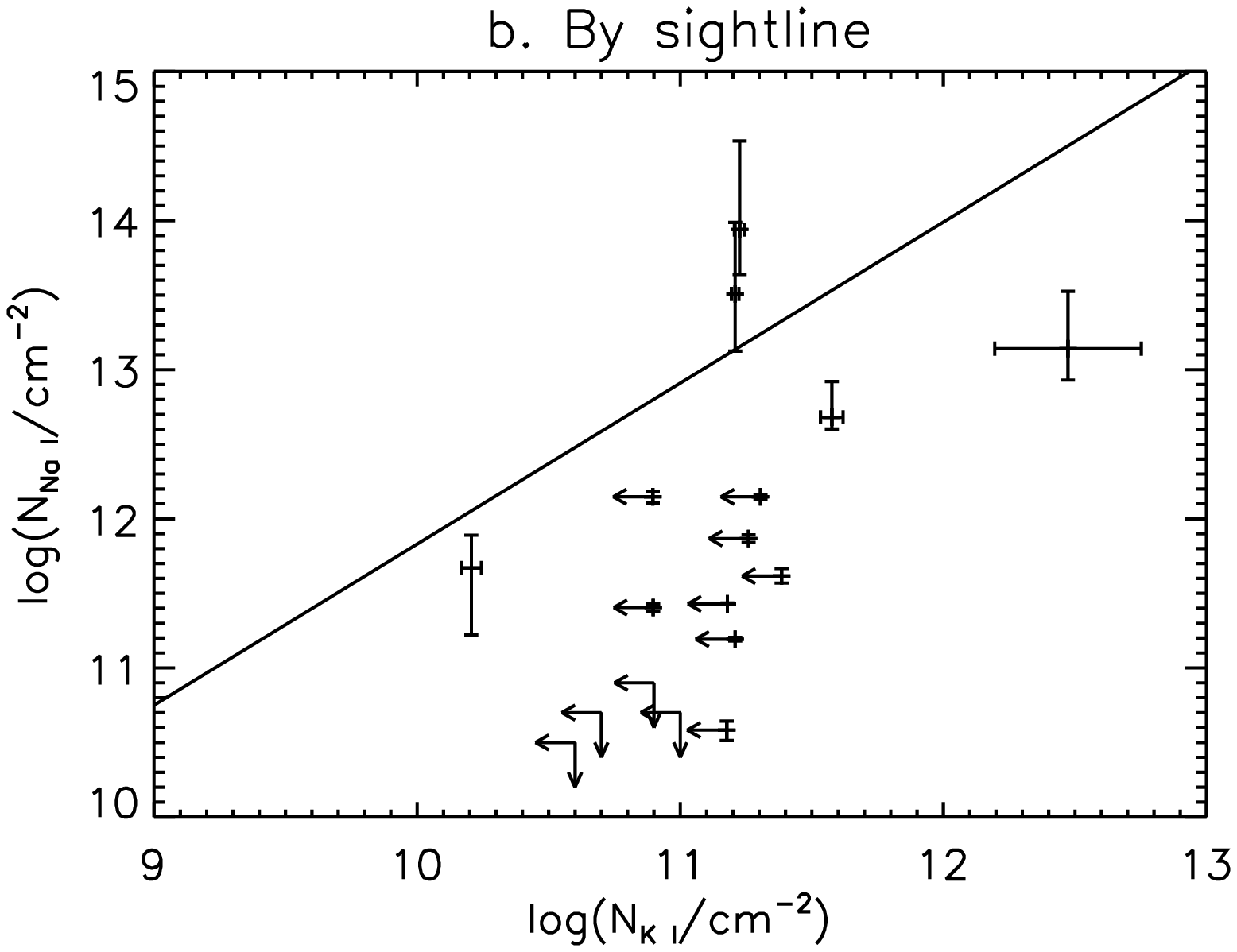}
\caption{Comparison of the column density in Na~\textsc{i} and K~\textsc{i} for each component where Na~\textsc{i} is detected (a., top) and for each sightline (b., bottom). Some error bars are smaller than the plot points. Upper limits in Na~\textsc{i} or K~\textsc{i} are shown as arrows. The line denotes the relationship between the column densities of these two ions found by \cite{WeltyHobbs01}. \label{na1k1comp}}
\end{figure}

Fig.\ \ref{na1k1comp} compares the Na~\textsc{i} and K~\textsc{i} column densities for the targets where both are detected, along with the relation between the respective column densities found by \cite{WeltyHobbs01}. The scatter around their relation is large, though larger when measured by component (Fig.~\ref{na1k1comp}a) than by sightline as a whole (Fig.~\ref{na1k1comp}b). We calculated $3\sigma$ upper limits on the K~\textsc{i} column density that could be associated with each Na~\textsc{i} component, under the assumption that $b_{\mathrm{K~\textsc{i}}}=b_{\mathrm{Na~\textsc{i}}}$. As can be seen in Fig.~\ref{na1k1comp}, these upper limits are largely uninformative.

Observations of two or more insterstellar species can be used to probe the temperature of and turbulence in the ISM through the Doppler parameter (line width).  From \cite{RedfieldT}, the relationship is
\begin{equation}
\label{Trelation}
b^2=\frac{2kT}{m}+\xi^2=0.016629\frac{T}{A}+\xi^2
\end{equation}
where $k$ is Boltzmann's constant, $m$ is the mass of the given ion, $\xi$ is the turbulent velocity, and $A$ is the mass of the ion in atomic mass units; the last part of Eqn.\ \ref{Trelation} assumes velocities in km s$^{-1}$.

\begin{figure}
\plotone{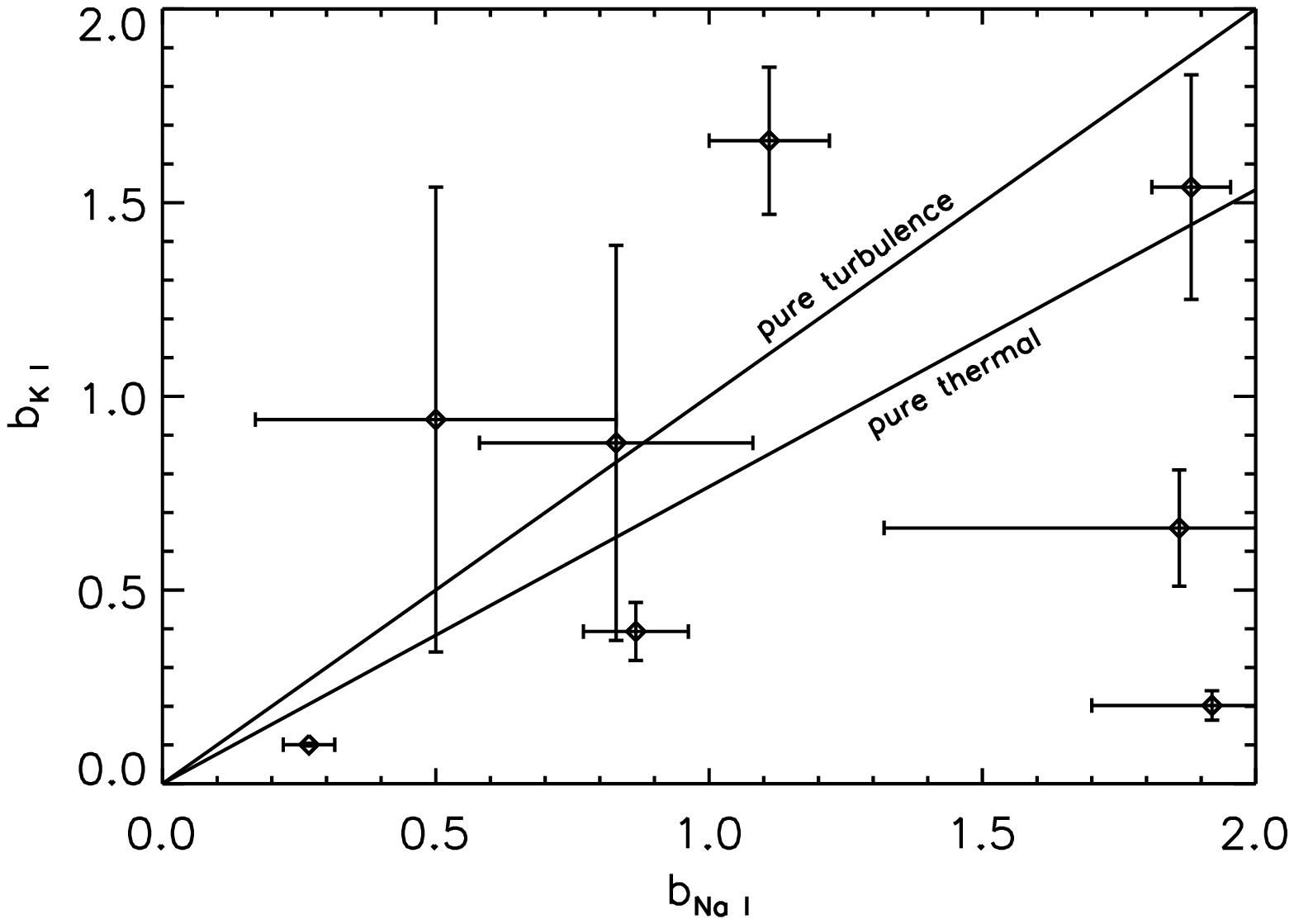}
\caption{Relationship between the Doppler parameters for Na~\textsc{i} and K~\textsc{i} for our data (points) and the theoretically expected range (between the lines correspondings to purely turbulent broadening, top, and purely thermal broadening, bottom). Some error bars are smaller than the plot symbols. \label{badtcomp}}
\end{figure}

From Eqn.\ \ref{Trelation} it is apparent that, for a given $b_{\mathrm{Na~\textsc{i}}}$, there is a range of physically allowed values of $b_{\mathrm{K~\textsc{i}}}$, with the maximum corresponding to pure turbulant broadening ($b_{\mathrm{K~\textsc{i}}}=b_{\mathrm{Na~\textsc{i}}}$) and the minimum corresponding to pure thermal broadening ($b_{\mathrm{K~\textsc{i}}}=\sqrt{A_{\mathrm{Na~\textsc{i}}}/A_{\mathrm{K~\textsc{i}}}}b_{\mathrm{Na~\textsc{i}}}$). These limits are shown in Fig.\ \ref{badtcomp}.

Of our eight components with both detected Na~\textsc{i} and K~\textsc{i}, only one lies within the physically allowed region in Fig.\ \ref{badtcomp} and two others are consistent with this region to within $1\sigma$. The remaining five components are more discrepant. This could occur if the Na~\textsc{i} and K~\textsc{i} are in general not well-mixed, i.e., they do not share the same temperature and/or turbulent velocity distribution. Nonetheless, we note that all of our detected K~\textsc{i} components lie within $1.2$ km s$^{-1}$ of a Na~\textsc{i} component, indicating that some physical relationship exists between the two ions, even if they are not well-mixed. Alternatively, as noted by \cite{WeltyHobbs01}, due to the differences in oscillator strength and abundance for Na~\textsc{i} and K~\textsc{i}, typically either the Na~\textsc{i} lines will be saturated or the corresponding K~\textsc{i} line will be very weak. Thus, our results could also be explained if we have underestimated the systematic errors on the Doppler parameters due to the saturation of the Na~\textsc{i} lines or the weakness of the K~\textsc{i} lines. There is no correlation between the ratio of the $b$ values and the distance to the target (as might be expected if blending of multiple clouds in longer sightlines results in corrupted line widths). There is also a large range in the ratio of the $b$ values among the components attributed to each of Clouds II and IV.

\subsection{Comparison with Previous Work}
\label{comparison}

To our knowledge, only two of our target stars have been previously observed in either Na~\textsc{i} or K~\textsc{i}. HIP 97165 was observed by \cite{Welty94} in Na~\textsc{i}; our results are consistent with theirs. \cite{ChaffeeWhite} observed both lines of the K~\textsc{i} doublet in absorption towards HIP 97757, albeit at lower resolution than the present work. They detected components at heliocentric velocities of $-25.0$ and $-11.4$ km s$^{-1}$, (or LSR velocities of $-10.1$ and $3.5$ km s$^{-1}$), likely corresponding to our components at $-8.473$ and $4.663$ km s$^{-1}$, respectively. Their Doppler parameters ($1.3_{-0.6}^{+1.0}$ km s$^{-1}$ and $0.8\pm0.1$ km s$^{-1}$, respectively) are somewhat larger than ours. Their measured logarithmic column densities ($11.26\pm0.14$ and $11.44_{-0.39}^{+0.06}$, respectively), deviate from ours by 3.2$\sigma$ and 1.2$\sigma$, respectively. These discrepancies likely result from their lower spectral resolution.
They did not detect our third component at a LSR velocity of $-4.90$ km s$^{-1}$; with $\log N_{\mathrm{K~I}}=10.420$, it was likely below their detection limit.

\cite{Lallement} presented maps of the neutral gas ISM density within $\sim$250 pc derived from Na~\textsc{i} equivalent width measurements and column densities. In the two maps in their work bracketing the \emph{Kepler} field (see their Figs.\ 7 and 8), a small cloud is visible at a distance of $\sim$ 50 pc. They identify this cloud with the Na~\textsc{i} component identified by \cite{Welty94} towards HIP 97165, and is thus the same as our Cloud II. They also identified absorption from this cloud towards HD 192640 (29 Cyg) and HD 193369 (36 Cyg), both located well outside the {\it Kepler} field of view, and quoted a heliocentric velocity of $-19$ km s$^{-1}$ for these components (corresponding to LSR velocities of $\sim -4$ km s$^{-1}$), in agreement with our values for Cloud II (between $\sim -5$ and $\sim-1$ km s$^{-1}$). This suggests an angular size for the cloud of $19.3^{\circ}$ and a physical size of 17.7 pc at a distance of 50.6 pc. However, \cite{Lallement} also noted that HD 192640 is a $\lambda$ Boo star, and so the Na~\textsc{i} absorption could be circumstellar in nature.

\cite{Welty96} also observed HIP 97165, albeit in Ca \textsc{ii}. They detected three absorption components, the weakest of which is coincident in velocity with our one detected Na~\textsc{i} component. \cite{RedfieldLinsky08} attributed this component to the Aql cloud, one of the LISM clouds close to the Sun. For other sightlines through the Aql cloud, they measured H~\textsc{i} column densities of $\log(N_{\mathrm{H~I}})=17.1-18.1$. Assuming that the H$_2$ fraction is negligible (reasonable for the modest density, hot conditions of the LISM), the relation between Na~\textsc{i} and H column density found by \cite{WeltyHobbs01} (see \S\ref{voldens} for more detail) predicts that the Na~\textsc{i} column density towards HIP 97165 due to the Aql cloud should be $\log (N_{\mathrm{Na~I}})<7.3$, well below our detection limits. We note that the data used to derive the \cite{WeltyHobbs01} relation do not extend below $\log (N_{\mathrm{H~I+H}_2})\sim19$, and so this estimate should be considered very uncertain; nonetheless, it suggests that our detection of Na~\textsc{i} towards HIP 97165 is not due to the Aql cloud, but rather due to the more distant, higher column density cloud detected by \cite{Lallement}, which is coincident in velocity with the Aql cloud. 

\subsection{Estimating Volume Densities}
\label{voldens}

\cite{WeltyHobbs01} found a relationship between Na~\textsc{i} and total H column density, namely (from Fig.~18 and associated text in that work),
\begin{equation}
\log N_{\mathrm{H~\textsc{i}}+\mathrm{H}_2}=0.478 \log N_{\mathrm{Na~\textsc{i}}}+14.6
\end{equation}
which we adopt to calculate the total H column density from our Na~\textsc{i} measurements. We note, however, that \cite{WeltyHobbs01} also found that even at a fixed Na~\textsc{i} column density, the total H column density can vary by up to $\sim1$ dex. Thus, our estimated column densities will not be particularly accurate.

In order to produce physically motivated upper and lower limits on the number density of each cloud, we need to have physically motivated limits on the cloud size along the line of sight, which, when coupled with the observed column densities, will yield limits on the number densities. We emphasize that the limits on cloud size that we derive here are used \emph{only} for computing the limits on number density, not for our later discussion of which planetary systems might lie within each cloud. Observed ISM clouds typically are in the form of sheets or filaments \citep[such as the interstellar cirrus discovered by IRAS;][]{Low84}, or, sometimes, rounder, more ``potato''-like shapes \citep[e.g., the LIC that currently surrounds our solar system;][]{RedfieldLinsky00}. Thus, we can obtain a reasonable upper limit on the size of a given cloud (and a lower limit on the number density) by assuming that it is spherical, i.e., that the depth of the cloud in the radial direction is equal to its maximum extent in the tangential direction. We estimate this by calculating the separation between the two most distant stars for which absorption is detected for a given cloud, and computing the corresponding transverse size at the distance corresponding to the distance upper limit for that cloud. We also need a lower limit on the cloud size in order to calculate an upper limit on the number density. \cite{Peek11} found that the cold, dense Local Leo Cold Cloud (LLCC), an island of cold neutral medium within the Local Bubble, has a width in the plane of the sky of $\sim0.25-0.54$ pc; assuming that its thickness along the line of sight is similar to its width (i.e., that the cloud is tubular), they estimated a number density of 150-320 cm$^{-3}$ for the cloud. Dense clouds and cores in star-forming regions can have thicknesses from $\sim0.15$ pc \citep{Lee14} down to 0.01 pc \citep{White15}. However, these clouds are very dense---\cite{Lee14} found densities of $\sim1-2\times10^5$ cm$^{-3}$, while smaller clouds will be even denser. Even with a thickness of 0.1 pc, our highest column density clouds would only have number densities of $\sim5\times10^3$ cm$^{-3}$, while if they have thicknesses of 0.5 pc they would have densities of the same order of magnitude as the LLCC. We thus conclude that thicknesses of 0.5 pc are more plausible for the types of clouds that we have observed.
 Moreover, such dense clouds tend to be physically small, making it unlikely that we would see such a cloud stretching over a significant portion of the {\it Kepler} field of view. We thus adopt 0.5 pc as a reasonable lower limit to the cloud thickness; the exact value for the lower limit is not overly important, as later we will only use the cloud density lower limits derived from the cloud size upper limits, not the density upper limits derived from the size lower limits, to estimate astrosphere sizes (see \S\ref{aspheresizes}). Note that this says nothing about the orientation of the cloud with respect to the line of sight; such a cloud could be oriented at an angle to the line of sight, and so have considerable overall depth (i.e., difference between the distances to the closest and farthest elements of the cloud) even if any given line of sight only transverses the cloud for 0.5 pc.

We then combine these size estimates with the maximum and minimum measured column densities for a given cloud to compute reasonable upper and lower limits for the number density in each cloud. In most cases these limits are rather unconstrained, but are consistent with measured volume densities in the ISM (see Table \ref{meantable} for the complete list of upper and lower limits). The upper limit for Clouds II is $\sim10^3$ cm$^{-3}$, likely an overestimate as this density would correspond to a molecular cloud, and the survey of \cite{Dame01} did not find any significant CO emission in the direction of the {\it Kepler} field.

For all six clouds, the minimum number densities that we derive are equal to or greater than that inferred for the LIC, the cloud currently surrounding our solar system. Thus, any planetary systems around solar-type stars that currently reside within these clouds can be expected to possess astrospheres smaller than the current heliosphere (for a similar velocity differential between the star and its exoLISM and similar interstellar magnetic field and ionization properties).

\begin{deluxetable*}{cccccccc}
\tablewidth{0pt}
\tablecaption{Range of Na~\textsc{i} Cloud Properties \label{meantable}}
\tablehead{ 
\colhead{Cloud Name} & \colhead{$N_{\mathrm{obs}}$}& \colhead{$v_{\mathrm{LSR}}$ (km s$^{-1}$)} & \colhead{$b$ (km s$^{-1}$)} & \colhead{$\log(N_{\mathrm{Na~I}}/\mathrm{cm}^{-2})$} & \colhead{$N_{\mathrm{Na~I}}/N_{\mathrm{K~I}}$} & \colhead{$n_{\mathrm{H}}$ (cm$^{-3}$)} & \colhead{Distance (pc)} \\
\colhead{(1)} & \colhead{(2)} & \colhead{(3)} & \colhead{(4)} & \colhead{(5)} & \colhead{(6)} & \colhead{(7)} & \colhead{(8)}
}

\startdata
I & 3 & $-8.63- -6.639$ & $0.249-4.46$  & $10.909-12.53$ & $1.4$ & $0.51-260$ & $62.5-340$ \\
II & 11 & $-4.94- -1.73$ & $0.268-2.9$ & $10.789-13.84$ & $29-2800$ & $1.5-1100$ & $<50.6$ \\
III & 9 & $0.81-1.42$ & $0.231-2.94$ & $10.583-12.496$ & \ldots & $0.58-240$ & $<129$ \\
IV & 5 & $3.65-5.369$ & $0.378-1.86$ & $11.664-13.45$ & $10-400$ & $0.73-690$ & $200-292$ \\
V & 2 & $7.348-8.194$ & $0.94-1.119$ & $11.902-11.997$ & \ldots & $1.2-140$ & $200-426$ \\
VI & 4 & $9.63-11.91$ & $0.298-1.90$ & $10.503-12.190$ & \ldots & $0.21-170$ & $200-426$ 
\enddata
\tablecomments{(1) Designation of cloud (see \S \ref{identification}). (2) Number of sightlines along which the cloud is detected in Na~\textsc{i}. (3) Range of observed LSR velocities for this cloud. (4) Range of Doppler parameters for this cloud. (5) Range of Na~\textsc{i} column densities for this cloud. (6) Range of Na~\textsc{i} to K~\textsc{i} column density ratios (i.e., abundance ratios) for this cloud. No detected K~\textsc{i} components were associated with Clouds I, III, or V.  (7) Range of possible hydrogen number densities for this cloud (see \S \ref{voldens}). (8) Range of possible distances to the cloud.} 
\end{deluxetable*}

\section{Discussion}

\subsection{Placement of Planets in Clouds}

Many of the confirmed \emph{Kepler} planets do not have reliable published distances, due in no small part to the faintness of the host star population on average. A handful, however, do have distances determined through a variety of methods. In Table \ref{planettable} we summarize all confirmed or validated planets in the \emph{Kepler} field of view with known distances of less than 450 pc. We choose this distance limit as 426 pc is the upper limit for the distance to Clouds V and VI, the most distant of our clouds observed on more than one sightline.

In addition to the \emph{Kepler} planets, there is one known and one suggested planet from radial velocity surveys in the \emph{Kepler} field of view. 16 Cyg B~b \citep{16CygBb} is an $m\sin i=1.68 M_J$ planet on an ${\sim800}$ day orbit located at a distance of 21.21 pc \citep{vanleeuwen07}. This is sufficiently close that it likely resides well within the Local Bubble \citep{Lallement}, and likely experiences a modest density interstellar environment similar to that of the Sun.

The closest of our target stars, HIP 96441 (a.k.a.\ $\theta$~Cyg), has been suggested to harbor one or more planets, but an unusual correlation between the bisector velocity span and the radial velocities makes this interpretation ambiguous \citep{thetaCygb}. Located at a distance of 18.34 pc, $\theta$ Cyg is likely also not surrounded by any of the ISM clouds detected in this sample, by the same reasoning used for 16 Cyg. Indeed, we do not detect any absorption towards this target.

Most of the nearby {\it Kepler} planets are on orbits close to their stars and thus are not in the habitable zone. There are, however, a few interesting systems in this sample. Most interesting for our purposes are Kepler-186, host of the first confirmed approximately Earth-size habitable zone planet, Kepler-186 f \citep{Kep186}, Kepler-22, host of the first confirmed \emph{Kepler} planet in the habitable zone, Kepler-22 b \citep{Kep22}, and several systems with small habitable zone planets statistically validated by \cite{Torres15}: Kepler-438, 296, 440, 441, 442, 437, and KOI-4427. We choose to include KOI-4427 in our sample, even though the planet candidate KOI-4427.01 has only been statistically validated through the exclusion of false positive scenarios to a confidence level of 99.2\%, unlike the other systems from \cite{Torres15}, which have been validated to a confidence level of 99.5\% or higher. Other systems include Kepler-16, host to the first confirmed transiting circumbinary planet \citep{Kep16}; Kepler-20, a system with five transiting super-earths and neptunes \citep{Kep20}; Kepler-42, a system of three sub-Earth-size planets in very close orbits around an M dwarf \citep{KOI961}; Kepler-37, host to three planets, including the smallest known transiting planet, with a radius only slightly larger than that of the Moon \citep{Kep37}; Kepler-421, host to the longest-period confirmed transiting planet, with an orbital period of 704 days \citep{Kep421}; and Kepler-444, an $\sim$11 Gyr old thick disk star with five transiting planets. In total, we consider 31 systems.

For each of these systems, we considered their distances and locations in the \emph{Kepler} field of view relative to our identified clouds to determine whether any of these planetary systems could be located within one of these clouds. We find that 13 of these systems could lie within one of our identified clouds. Many of the \emph{Kepler} systems, however, are located in the central or northwestern parts of the \emph{Kepler} field of view, where we do not have any target stars at distances similar to these exoplanet hosts. Thus, our current sampling is insufficient to determine whether these systems could be located inside our ISM clouds. 
See Table \ref{planettable} for a summary. 

\subsection{Estimating Astrosphere Sizes}
\label{aspheresizes}

Given this discussion regarding the placement of {\it Kepler} systems within our ISM clouds, we wish to estimate plausible astrosphere sizes. The size of an astrosphere, however, depends not only upon the exoLISM density and velocity but also on the parameters of the outflowing stellar wind. The stellar wind parameters are very difficult to observe for solar-type stars, due to the small mass flux. \cite{Wood05}, however, measured the mass-loss rate $\dot{M}$ for several nearby stars and provided a scaling relation between $\dot{M}$ and the x-ray flux $F_X$ (for $F_X<8\times10^5$ ergs cm$^{-2}$ s$^{-1}$). If we had x-ray fluxes for the \emph{Kepler} planet-host stars, we could thus infer $\dot{M}$.

We searched for x-ray emission from these stars by querying the NEXXUS 2 database\footnote{http://www.hs.uni-hamburg.de/DE/For/Gal/Xgroup/nexxus/nexxus.html} at the position of each of the host stars listed in Table \ref{planettable}. Unfortunately, none of these stars has been detected in x-rays. It would thus be helpful to obtain x-ray fluxes for \emph{Kepler} planet-host stars in order to perform a more rigorous version of this analysis. Instead, we will use scaling relations based on the stellar mass and age in order to estimate the astrosphere sizes. Ideally, the astrosphere sizes would be modeled using a detailed multifluid model, such as that used by \cite{Muller06} to model the response of the heliosphere to differing LISM conditions; however, given the current uncertainty in the parameters of exoplanet host stars and their ambient ISM conditions, such a detailed approach is not warranted at this time.

From \cite{SmithScalo}, the size of a pressure-supported astrosphere moving supersonically and superalfvenically through the ISM, so that the dominant source of external pressure is ram pressure, is
\begin{equation}
\frac{r_a}{r_0}=\bigg(\frac{n_0}{n}\bigg)^{1/2}\frac{v_0}{V}
\end{equation}
where $r_a$ is the astrosphere radius, $r_0$ is an arbitrary reference radius within the astrosphere, $n_0$ and $v_0$ are the stellar wind number density and velocity, respectively, at $r_0$, $n$ is the ISM number density, and $V$ is the ISM streaming velocity.
\cite{SmithScalo} parameterized the dependence of the stellar wind on the age and mass of the star in a function
\begin{equation}
\xi(M_*,t_*)=\bigg(\frac{M_*}{M_{\odot}}\bigg)^{\alpha}\bigg(\frac{t_*}{\mathrm{4.5 Gyr}}\bigg)^{\beta}
\end{equation}
where $M_*$ and $t_*$ are the stellar mass and age, respectively. Thus, they had
\begin{equation}
\frac{r_a}{r_0}=\bigg(\xi(M_*,t_*)\frac{n_0}{n}\bigg)^{1/2}\frac{v_0}{V}
\end{equation}
where $r_0$, $n_0$, and $v_0$ are now appropriate values for the Sun. We can write the solar wind density $n_0$ in terms of the solar mass-loss rate, and thus obtain an absolute astrosphere size
\begin{equation}
r_a=\frac{1}{V}\bigg(\xi(M_*,t_*)\frac{v_0\dot{M}_{\odot}}{4\pi n m}\bigg)^{1/2}
\end{equation}
where $m$ is the mean mass of a solar wind atom (which we assume to be equal to the mass of a hydrogen atom). This assumes that $v_0$ is constant over mass and age of the stars, which is a typical assumption of previous works \citep{SmithScalo,Wood05}. As discussed by \cite{Wood04}, this is probably a not unreasonable assumption given that the surface gravity of most solar-type main sequence stars is similar, but the wind velocity may be higher for rapidly rotating stars.

\cite{SmithScalo} noted that the values of the power law indices $\alpha$ and $\beta$ are very uncertain. Using data from \cite{Wood05} and \cite{PenzMicela08} they calculated $\beta=-2.33$ and $\beta=-1.8$, respectively. We therefore adopt $\beta=-2$ for our calculations. \cite{SmithScalo} also noted that the value of $\alpha$ is even more uncertain; observations by \cite{Wood05} suggested that lower-mass and typically more active stars have stronger winds (i.e., $\alpha<0$). Once the x-ray flux increases beyond $\sim10^6$ erg cm$^{-2}$ s$^{-1}$, however, a dramatic drop in the mass loss rate is observed for these stars \cite{Wood14}, resulting in a much weaker wind (i.e., $\alpha>0$). With a lack of concrete information, we adopt $\alpha=0$. For comparison we also calculated astrosphere sizes for $\alpha=-1,1$; for sun-like stars the differences are negligible, but for the lowest-mass stars this results in an uncertainty of a factor of 10 on the astrosphere size. This highlights the need for reliable x-ray flux measurements of planetary host stars, particularly for the lowest mass stars.

For each star, we adopt parameters for the literature for $M_*$ and $t_*$. For the velocity $V$ of the star with respect to the ISM, we obtain the radial velocity of the star from the literature, convert this velocity to the LSR frame, compute $\Delta v=|v_*-v_{\mathrm{cloud}}|$, and finally assume that $V=\sqrt{3}\Delta v$, as the full three-dimensional space velocity should be, on average, $\sqrt{3}$ larger than the line-of-sight velocity.

In order to calibrate the \cite{SmithScalo} model, we tested it against the 27 multifluid heliosphere models of \cite{Muller06} by taking the relevant initial conditions for each of these models ($n_{\mathrm{ISM}}$, $v_{\mathrm{ISM}}$) and using them as inputs for the \cite{SmithScalo} model. For the \cite{Muller06} Model 1 we overpredict the heliosphere radius by a factor of 3.6; however, this is unsurprising as the initial conditions for Model 1 are a very hot, rarefied ISM, where the ISM pressure is dominated by thermal pressure, whereas the \cite{SmithScalo} model assumes that ram pressure is the dominant source of external pressure. For the other 26 models, the discrepancies between the two models are much smaller; the \cite{SmithScalo} model underpredicts the \cite{Muller06} astrosphere radii by a factor of 1.14 to 2.18, with a mean of 1.54 and a standard deviation of 0.24. We thus choose to make an empirical correction to the \cite{SmithScalo} model to allow it to better match the \cite{Muller06} model results, by multiplying the \cite{SmithScalo} astrosphere radii by a factor of 1.54. This corrected \cite{SmithScalo} model reproduces Models 2-27 of \cite{Muller06} with a standard deviation of 0.16.

We also tested this empirically corrected model on the heliosphere, using the density of the LIC from \cite{SlavinFrisch08} and the relative velocity between the Sun and the LIC from \cite{McComas12}. This yielded a heliosphere size of 154 AU, rather larger than the heliopause distance of 121 AU measured by Voyager 1. Our calculations are for the heliopause distance in the upwind direction, and since Voyager 1 is not moving in this direction, the actual heliopause size with which we should compare our model will be $<121$ AU.
This suggests that our model is accurate to no better than $30$\%. Given the uncertainty in the hydrogen densities calculated from the relation of \cite{WeltyHobbs01}, however, this level of inaccuracy in the astrosphere model will be only a small contribution to our error budget.

We perform this calculation for each of the {\it Kepler} planets which we find could be located within an ISM cloud. As noted earlier, we consider our upper limits on hydrogen number density for most of our clouds to be much higher than reasonable values, and so for these calculations we use our lower limits on $n_H$ for each cloud, resulting in upper limits on astrosphere size. These estimates are presented in Table~\ref{planettable}. We were unable to locate measurements of the absolute radial velocities of 
 HAT-P-11 (Kepler-3) and Kepler-45 in the literature, and so do not estimate astrosphere sizes for these systems. We also note that the value of $\beta=-2$ gives a strong dependence on the stellar age; as accurately measuring stellar ages is notoriously difficult, these estimates should be treated with caution. For example, there are two somewhat conflicting ages in the literature for Kepler-20. \cite{Kep20} obtained an age of $8.8^{+4.7}_{-2.7}$ Gyr from an isochrone analysis using the stellar mean density measured using the transit lightcurve. \cite{WalkowiczBasri13} calculated an age of 4.0 Gyr using gyrochronology, where they measured the stellar rotation period using spot modulation in the {\it Kepler} lightcurve and used this and the relationship between stellar age and rotation rate to measure the stellar age. We adopt the isochrone age, as \cite{WalkowiczBasri13} noted that the gyrochronological age-rotational period relations are significantly uncertain for stars of similar or greater age than the Sun; these relations are typically calibrated using open clusters of known age, and old open clusters are very rare. We note that the gyrochronological age for Kepler-20 is still compatible with the isochrone age to within $2\sigma$. Using the gyrochronological age rather than the isochrone age gives astrosphere sizes larger by a factor of $\sim2$. None of the upper limits on astrosphere sizes are small enough to put the known planets outside of the astrosphere; estimates range from 2.9 AU for Kepler-444 to 530 AU for Kepler-21. 
While we calculate an upper limit of 530 AU for Kepler-21, this may not be an accurate estimate as the star is a subgiant \citep[spectral type F5IV;][]{Kep21}. The relationship between stellar x-ray flux and mass loss found by \cite{Wood05} upon which our estimates ultimately rests was derived for main sequence stars, and none of the three subgiants in the sample of \cite{Wood05} agree with the relationship found for main sequence stars (although none of the three have x-ray fluxes within the range for which this relation is a good fit to the data). Additionally, as noted above, rapidly rotating stars may have higher stellar wind velocities than slowly rotating stars like the Sun, which could affect our estimated astrosphere sizes for Kepler-448 \citep[$v\sin i=60$ km s$^{-1}$;][]{Kep448}.

The remainder of this discussion is conditional upon the {\it Kepler} targets that we have investigated actually lying within the ISM clouds that we have identified; if the stars do not lie within these clouds, then their astrospheres will likely be larger than we have estimated. We note again that we only calculate upper limits on astrosphere size, conditional upon this assumption. We find that Kepler-444, 42, 445, and 20 could have astrospheres significantly smaller than the present-day heliosphere; Kepler-32 and 88 (the latter if in Cloud V) could have astrospheres similar in size to the present-day heliosphere, or smaller; and Kepler-88 (if in Cloud VI) could have an astrosphere much larger than the present-day heliosphere. 
Among the habitable zone planet hosts in our sample, Kepler-186 has an estimated maximum astrosphere size of 59 AU, somewhat smaller than the present-day heliosphere, while Kepler-437 and KOI-4427 could host astrospheres much larger than the heliosphere.

Another system of note is Kepler-444, the first {\it Kepler} planet host that is a member of the thick disk \citep{Kep444}. As a member of the thick disk, Kepler-444 has a large peculiar velocity of 154 km s$^{-1}$ and an old age of 11.2 Gyr. Due to this large velocity and old age, we calculate maximum astrosphere sizes of 2.9 and 4.5 AU if it should lie within Cloud II or III, respectively. This is still well outside the planetary orbits in this very compact system; the outermost of the five known planets, Kepler-444 f, has an orbital semi-major axis of just 0.0811 AU. We note that due to the systematically large velocities and old ages of the thick disk and halo populations, \citep[halo stars can also host planets; Kapteyn's Star may have a planet:][]{AngladaEscude14,Robertson15}, 
 such stars should have systematically smaller astrospheres than thin disk stars. As a result any planets around such stars will be more easily descreened by passage through interstellar clouds. Due to their high velocities, however, passages through such clouds will be shorter than for thin disk stars.

\subsection{Future Prospects}

The distances to most \emph{Kepler} planet candidate host stars are very uncertain or unknown, as these objects are too faint to have been observed by \emph{Hipparcos}. Although \emph{Gaia} will provide distances to the \emph{Kepler} hosts, until these data are released the ISM clouds in the {\it Kepler} field of view can be used as distance markers. Given our distance limits on the ISM clouds in the {\it Kepler} field of view, if absorption at a given velocity is observed in the spectrum of a target of interest, this target must be at least as distant as the intervening cloud. 

This work will also have bearing on a different distance indicator. \cite{SilvaAguirre12} demonstrated that they can derive distances from asteroseismology to accuracies of 5\%, but that reddening to the target star is the major source of error for more distant targets. Our work to map the ISM, the source of reddening, in the \emph{Kepler} field of view will inform the reddening corrections used to obtain these distances.

\emph{Gaia} will have additional bearing on future work on this field, as it will provide more precise distances to the \emph{Hipparcos} stars that we have observed, as well as distances to many fainter, more distant early-type stars in the \emph{Kepler} field of view. This will allow us to compile a map of the ISM in the \emph{Kepler} field of view with both higher spatial resolution and extending to greater distances. Additionally, it will provide distances to the {\it Kepler} planet host stars, allowing us to identify a larger sample of planet-hosting stars that might lie within ISM clouds.

Further observations are capable of constructing a more densely sampled map of the ISM in the {\it Kepler} field than we present here. The {\it Hipparcos} re-reduction of \cite{vanleeuwen07} contains more than 300 stars meeting our target selection criteria (but discarding our magnitude limits) that lie in or near the {\it Kepler} field of view, so there is no lack of potential targets.

\section{Summary and Conclusions}

The interstellar environments of planets, especially habitable planets, can have an impact upon the planetary habitability and climate. This occurs as the astrosphere size is regulated by the surrounding ISM (the exoLISM) density and streaming velocity, and in turn regulates the cosmic ray flux experienced by the planet \citep[e.g.,][]{Muller06}. Additionally, interstellar gas and dust can be deposited directly onto the planet. These effects can alter the planetary cloud cover, surface temperatures, ozone layer, surface ultraviolet radiation flux, and more \citep[e.g.,][]{YeghikyanFahr04,Pavlov05}. Thus, as future efforts are made to establish the climate and habitability of habitable zone planets, it is important to also consider the ambient interstellar environment.

We have presented the results of a small survey of the ISM within the {\it Kepler} prime mission field of view, the first focused on this region of the sky. We have measured the Na~\textsc{i} and K~\textsc{i} absorption towards a sample of early-type stars. Using these data we have identified six clouds located at distances of less than 450 pc, which we designate Clouds I through VI. We have found that Cloud II must lie at a distance of less than 56 pc, placing it firmly inside the Local Bubble. All six clouds likely have volume densities greater than that of the LIC. 
 In addition, we have identified five velocity components which are detected along only one line of sight.

Using the constraints on these ISM clouds, we have identified {\it Kepler} systems with confirmed planets which could lie within these clouds. Using the estimated cloud parameters, we have then estimated maximum astrosphere sizes for these systems, conditional upon these systems actually lying within these clouds. Most interestingly, we find that the astrosphere surrounding the habitable zone planet Kepler-186 f could be smaller than that of the Sun ($r_{a,\mathrm{max}}=59$ AU), while the thick disk star and planet host Kepler-444 could have an astrosphere just a few AU in size. Additionally, several known multiplanet systems (i.e. Kepler-20, Kepler-42, and Kepler-445) could have astrospheres much smaller than the present-day heliosphere (astropause distances of a few tens of AU), while the habitable zone planet hosts Kepler-437 and KOI-4427 could have astrospheres much larger than the present-day heliosphere. We note again that these estimates may be significantly in error due to uncertainties in the cloud number densities and inaccuracies in the astrosphere model.

While our work has relatively low spatial sampling and we can only identify clouds with more than one absorption component out to a distance of 450 pc, future work to observe more early-type stars in the {\it Kepler} field of view can better constrain the ISM properties in this key region of exoplanetary interest. Such a map will also bear on a number of other astrophysical applications, such as reducing the uncertainties on asteroseismic investigations and constraining the distances to {\it Kepler} planet hosts.

\vspace{12pt}

Thanks to Katherine Wyman for assistance with the ISM fitting routines. Thanks to the staff at McDonald Observatory, in particular David Doss, for invaluable assistance during the observing run. M.C.J.\ is supported by a NASA Earth and Space Science Fellowship under Grant NNX12AL59H.

This paper includes data taken at The McDonald Observatory of The University of Texas at Austin.
This research has made use of the SIMBAD database and the VizieR catalog access tool, both operated at CDS, Strasbourg, France.

\clearpage
\begin{turnpage}
\begin{deluxetable}{ccccccccccccc}
\tablecolumns{11}
\tablecaption{Confirmed or Validated Planets within 450 pc \label{planettable}}
\tablehead{ 
\colhead{\emph{Kepler} ID} & \colhead{KOI} & \colhead{$T_\mathrm{eff}$} & \colhead{$Kp$} & \colhead{d (pc)} & \colhead{Clouds} & \colhead{$M_* (M_{\odot})$} & \colhead{age (Gyr)} & \colhead{$N_{\mathrm{planets}}$} & \colhead{$r_{a,\mathrm{max}}$ (AU)} & \colhead{Ref.}\\
\colhead{(1)} & \colhead{(2)} & \colhead{(3)} & \colhead{(4)} & \colhead{(5)} & \colhead{(6)} & \colhead{(7)} & \colhead{(8)} & \colhead{(9)} & \colhead{(10)} & \colhead{(11)} 
}

\startdata
\ldots & Sun & 5780 & -26.83 & $4.8\times10^{-6}$ & LIC & 1.0 & $4.6$ & 8 & 154, 121\tablenotemark{g} & 1 \\
\ldots & 16 Cyg B & $5772 \pm 25$ & 6.0 & 21.21 $\pm$ 0.12\tablenotemark{b} & \ldots & \ldots & \ldots & 1 & \ldots & 2\\
444 &  3158 & $5046 \pm 74$ &  8.717 & $35.7 \pm 1.1$\tablenotemark{b} &  II, III &  $0.758 \pm 0.043$ &  $11.23^{+0.91}_{-0.99}$ &  5 &  2.9, 4.5 &  3, 4 \\
3 & HAT-P-11 & $4850 \pm 50$ & 9.2 & 36.4 $\pm$ 1.3\tablenotemark{b} & II & $0.81^{+0.02}_{-0.03}$ & $6.5^{+5.9}_{-4.1}$ & 1 & \ldots & 5\\
42 & 961 & $3200 \pm 65$ & 15.92 & 38.7 $\pm$ 6.3\tablenotemark{c} & II & $0.13 \pm 0.05$ & $>4.5$ & 3 &  11 & 6\\
37 & 245 & $5417 \pm 75$ & 9.701 & $\sim66$\tablenotemark{d} & \ldots & \ldots & \ldots & 3 & \ldots & 7\\
445 &  2704 & $3157 \pm 60$ &  \ldots & $\sim90$\tablenotemark{c} & I &  $0.18 \pm 0.04$ &  $>5$ &  3 & 30 &  8\\
21 & 975 &  $6131 \pm 44$ & 8.2 & 108 $\pm$ 10\tablenotemark{b} & III & $1.340 \pm 0.010$ & $2.84 \pm 0.34$ & 1 & 530\tablenotemark{h} & 9\\
446 &  2842 & $3359 \pm 60$ &  \ldots & $\sim120$\tablenotemark{c} &  \ldots &  \ldots &  \ldots &  3 &  \ldots &  8\\
410A & 42 & $6325 \pm 75$ & 9.364 & $132 \pm 6.9$\tablenotemark{d} & \ldots & \ldots & \ldots & 2\tablenotemark{f} & \ldots & 10 \\
68 & 246 & $5793 \pm 74$ & 10.0 & $135 \pm 10$\tablenotemark{d} & \ldots & \ldots & \ldots & 3 & \ldots & 11\\
438  &  3284 & $3748 \pm 112$ &  14.467 & $145^{+20}_{-23}$\tablenotemark{c} &  \ldots &  \ldots &  \ldots &  1 &  \ldots &  12 \\
186 & 571 & $3788 \pm 54$ & 14.6 & $151 \pm 18$\tablenotemark{c} & I & $0.478 \pm 0.055$ & 2.55\tablenotemark{e} & 5 & 59 & 13\\
10 & 72 &  $5680 \pm 91$ & 10.96 & 173 $\pm$ 27\tablenotemark{d} & \ldots & \ldots & \ldots & 2 & \ldots & 14\\ 
22 & 87 &  $5518 \pm 44$ & 11.664 & 190 & \ldots & \ldots & \ldots & 1 & \ldots & 15\\
63 &  63 & $5576 \pm 50$ &  11.6 & $200\pm15$\tablenotemark{c} &  \ldots &  \ldots &  \ldots &  1 &  \ldots &  16 \\
16 & 1611 & $4450 \pm 150$\tablenotemark{a}  & 11.762 & $\sim$200\tablenotemark{c} & \ldots & \ldots & \ldots & 1 & \ldots & 17\\
296 &  1422 & $3572 \pm 80$ &  15.921 & $226^{+28}_{-18}$\tablenotemark{c} &  \ldots &  \ldots &  \ldots &  5 &  \ldots &  12 \\
1 & TrES-2 &  $5960 \pm 100$ & 11.3 & $\sim$230\tablenotemark{c} & \ldots & \ldots & \ldots & 1 & \ldots & 18 \\
\ldots &  4427 & $3813 \pm 112$ &  15.645 & $240^{+32}_{-39}$\tablenotemark{c} &  V, VI & $0.526^{+0.040}_{-0.062}$ & $3.6^{+2.6}_{-1.3}$ &  1 & 300, 380 &  12 \\
440 &  4087 & $4134 \pm 154$ &  15.134 & $261^{+16}_{-46}$\tablenotemark{c} &  \ldots &  \ldots &  \ldots &  1 &  \ldots &  12 \\
441 &  4622 & $4340 \pm 177$ &  15.142 & $284^{+28}_{-48}$\tablenotemark{c} &  \ldots &  \ldots &  \ldots &  1 &  \ldots &  12 \\
20 & 70 &  $5455 \pm 100$ & 12.498 & 290 $\pm$ 30\tablenotemark{c} & IV, VI & $0.912 \pm 0.034$ & $8.8^{+4.7}_{-2.7}$ & 5 & 54, 63 & 19 \\
32 & 952 & $3793^{+80}_{-74}$ & 15.913 & 303 $\pm$ 14\tablenotemark{c} & I, VI & $0.54 \pm 0.02$ & 3.21\tablenotemark{e} & 6 & 180, 99 & 20 \\ 
2 & HAT-P-7 & $6350 \pm 80$ & 10.5 & 320$^{+50}_{-40}$\tablenotemark{c} & \ldots & $1.47^{+0.08}_{-0.05}$ & $2.2 \pm 1.0$ & 1 & \ldots & 21\\
421 &  1274 & $5308 \pm 50$ &  13.354 & $320 \pm 20$\tablenotemark{c} &  \ldots &  \ldots &  \ldots &  1 &  \ldots &  22 \\
45 & 254 &  $3820 \pm 90$ & 15.979 & 333 $\pm$ 33\tablenotemark{c} & V, VI & $0.59 \pm 0.06$ & 0.69\tablenotemark{e} & 1 & \ldots & 23\\
88 & 142 & $5471 \pm 50$ & 13.113 & $339_{-23}^{+25}$\tablenotemark{c} & V, VI & $0.956_{-0.051}^{+0.041}$ & $2.2_{-2.0}^{+2.4}$ & 2\tablenotemark{f} & 140, 270 & 24, 25 \\
442 &  4742 & $4402 \pm 100$ &  14.976 & $342^{+19}_{-22}$\tablenotemark{c} &  \ldots &  \ldots &  \ldots &  1 &  \ldots &  12 \\
62 & 701 & $4925 \pm 70$ & 13.75 & 368\tablenotemark{c} & \ldots & \ldots & \ldots & 5 & \ldots & 26 \\
437 &  3255 & $4551 \pm 100$ &  14.352 & $417^{+24}_{-21}$\tablenotemark{c} & VI & $0.707_{-0.027}^{+0.033}$ & $2.9_{-0.3}^{+7.5}$ &  1 & 290 &  12 \\
448 & 12 & $6820 \pm 120$ & 11.353 & $426\pm40$\tablenotemark{c} & V, VI & $1.452 \pm 0.093$ & $1.5 \pm 0.5$ & 1 & 170, 340 & 27

\enddata

\tablecomments{Confirmed or validated nearby planetary systems within the {\it Kepler} field of view, along with the Sun for comparison. (1) \emph{Kepler} number. (2) KOI number, or alternate name for pre-\emph{Kepler} planets. (3) Effective temperature of the host star. (4) Host star magnitude in the \emph{Kepler} bandpass. (5) Distance in parsecs. (6) ISM clouds within which the system could lie; see text for more details. (7) Mass of the host star. (8) Age of the host star. (9) Number of known planets orbiting the star. (10) Estimated maximum astrosphere size. For stars which overlap with more than one cloud, the sizes calculated using the parameters of each cloud are listed in the same order as the clouds in column (6). (11) Reference for columns 3, 5, 7, and 8 (unless noted otherwise). A second reference denotes the source of the stellar radial velocity used in computing $r_{a,\mathrm{max}}$. In the interests of brevity, values are not given for columns 7 or 8 if the system cannot lie within one of the ISM clouds.\\
References. (1) \cite{Lang}; (2) \cite{16CygBb}; (3) \cite{Kep444}; \cite{Kep444vel}; (5) \cite{HATP11}; (6) \cite{KOI961}; (7) \cite{Kep37}; (8) \cite{Muirhead15}; (9) \cite{Kep21}; (10) \cite{Kep410A}; (11) \cite{Kep68}; (12) \cite{Torres15}; (13) \cite{Kep186}; (14) \cite{Kep10}; (15) \cite{Kep22}; (16) \cite{Kep63}; (17) \cite{Kep16}; (18) \cite{TrES2}; (19) \cite{Kep20}; (20) \cite{Kep32}; (21) \cite{HATP7}; (22) \cite{Kep421}; (23) \cite{Kep45}; (24) \cite{Kep88}; (25) \cite{Kep88vel}; (26) \cite{Kep62}; (27) \cite{Kep448}.}

\tablenotetext{a}{Circumbinary planet. The listed $T_{\mathrm{eff}}$ is that of the primary star; only the mass of the secondary star is known (0.20255 $M_{\odot}$).}
\tablenotetext{b}{Distance from \emph{Hipparcos} data.}
\tablenotetext{c}{Distance estimated from stellar properties and observed magnitude.}
\tablenotetext{d}{Distance determined from asteroseismic analysis.}
\tablenotetext{e}{Age estimated using gyrochronology by \cite{WalkowiczBasri13}.}
\tablenotetext{f}{One transiting planet plus an additional non-transiting companion detected via transit timing variations.}
\tablenotetext{g}{Estimated using our model (first figure; see text). As measured by Voyager 1 \citep[second figure; see][]{Gurnett13}.}
\tablenotetext{h}{The estimated maximum astrosphere size for Kepler-21 may be incorrect because the host star is a subgiant, not a main sequence star; see text for details.}

\end{deluxetable}

\end{turnpage}
\clearpage

\end{document}